\documentclass[pdflatex,sn-mathphys-ay,referee]{sn-jnl}

\usepackage{graphicx}%
\usepackage{multirow}%
\usepackage{amsmath,amssymb,amsfonts}%
\usepackage{amsthm}%
\usepackage{mathrsfs}%
\usepackage[title]{appendix}%
\usepackage[dvipsnames, table]{xcolor}%
\usepackage{textcomp}%
\usepackage{manyfoot}%
\usepackage{booktabs}%
\usepackage{algorithm}%
\usepackage{algorithmicx}%
\usepackage{algpseudocode}%
\usepackage{listings}%
\usepackage{makecell, pdflscape}
\usepackage{subcaption}
\usepackage{comment}
\usepackage{lineno} 

\theoremstyle{thmstyleone}%
\newtheorem{theorem}{Theorem}%
\theoremstyle{thmstyletwo}%

\theoremstyle{thmstylethree}%

\begin{document}

\title[Article Title]{Basic Reproductive Numbers in Host-Parasite Systems with Demographic Bistability}


\author*[1]{\fnm{Carli} \sur{Peterson}}\email{cxp4986@mavs.uta.edu}

\author[1]{\fnm{Erika} \sur{Gallo}}\email{erika.gallo@uta.edu}

\author[1,2]{\fnm{Christopher} \sur{Kribs}}\email{kribs@uta.edu}

\affil[1]{\orgdiv{Department of Mathematics}, \orgname{The University of Texas at Arlington}, \orgaddress{\street{411 S Nedderman Dr}, \city{Arlington}, \postcode{76019}, \state{TX}, \country{USA}}}

\affil[2]{\orgdiv{Department of Teacher and Administrator Preparation}, \orgname{The University of Texas at Arlington}, \orgaddress{\street{701 Planetarium Pl}, \city{Arlington}, \postcode{76019}, \state{TX}, \country{USA}}}

\abstract{The basic reproductive number, $\mathcal{R}_0$, is widely used to determine whether an infectious disease can persist in a population. However, its interpretation may become unclear when the underlying disease-free demographic dynamics exhibit multiple types of stable attractors. In this study, we investigate how demographic bistability between fixed points and periodic cycles affects the calculation and interpretation of $\mathcal{R}_0$. We consider two discrete-time host-parasite models with infection: a tick-host model motivated by \textit{Borrelia burgdorferi} transmission and a Nicholson-Bailey host-parasitoid model with intraspecific competition and infection. For each model, we calculate reproductive numbers associated with different disease-free attractors using the next-generation operator method and compare their invasion thresholds. In the tick-host model, the reproduction number calculated on the stable demographic cycle differs from that calculated at the positive coexistence fixed point, resulting in parameter regions where infection can invade at one demographic attractor but not the other. In contrast, in the host-parasitoid model, the host-driven invasion threshold is necessarily the same at the positive coexistence fixed point and on the host-only periodic cycles. These contrasting results demonstrate that demographic bistability alone does not necessarily produce attractor-dependent invasion thresholds. Rather, the effect of demographic dynamics on invasion depends on the structure of pathogen transmission. In particular, a complete host--vector--host transmission pathway can allow differences between demographic attractors to alter the invasion threshold. Our results emphasize the importance of considering both demographic dynamics and transmission structure when interpreting $\mathcal{R}_0$ in populations with multiple stable disease-free states.} 


\maketitle

\section{Introduction}

In mathematical epidemiology, the basic reproductive number, $\mathcal{R}_0$, is one of the most widely used metrics for characterizing pathogen invasion and persistence \citep{heffernan2005perspectives,diekmann2009construction}. Classically, $\mathcal{R}_0$ is defined as the expected number of secondary infections generated by a single infected individual introduced into an otherwise fully susceptible population. For systems with a unique disease-free fixed point, $\mathcal{R}_0<1$ implies local asymptotic stability of the disease-free state, while $\mathcal{R}_0>1$ indicates the potential for pathogen invasion. Common formulations of the next-generation operator are typically presented for systems in which the disease-free population approaches a unique asymptotically stable fixed point \citep{diekmann1990definition, roberts2007pluses, van2002reproduction}. However, nonlinear systems may exhibit multiple coexisting stable attractors under the same parameter set, including combinations of stable fixed points and periodic cycles. In such cases, the appropriate interpretation of $\mathcal{R}_0$ becomes less clear: if pathogen invasion is evaluated about different disease-free attractors, does the resulting invasion threshold depend on the attractor? 

Multistability has been documented in a variety of models, including both continuous-time \citep{jin2007sirs, juher2023saddle, bate2013complex, baer2006multiple, boerlijst2015competition, scarabel2025bistability} and discrete-time \citep{neubert1992subcritical, kon2006multiple, beddington1975dynamic, lopez2005indirect} systems. 
Discrete-time models provide a particularly natural framework for studying this question because seasonal, generational, and other periodic population processes can be represented through discrete updates, and the resulting nonlinear dynamics can exhibit stable periodic cycles. Although methods for computing reproductive numbers at both fixed points and periodic cycles have been well established \citep{van2019demographic, cushing2012net, caswell2009stage}, considerably less attention has been given to systems in which multiple disease-free attractors coexist under the same parameter set. In these cases, pathogen invasion can be evaluated about each attractor, potentially yielding distinct reproduction numbers and, consequently, different invasion thresholds. Thus, an important question is whether demographic bistability necessarily leads to attractor-dependent invasion thresholds or whether additional features of the epidemiological system are required.

This question is particularly relevant to vector-borne diseases because pathogen transmission depends on interactions between two populations: hosts and vectors. As a result, changes in the underlying demographic state can affect both components of the host--vector--host transmission pathway and may alter the conditions for pathogen invasion. 

To investigate this question, we consider two discrete-time models exhibiting bistability between a disease-free fixed point and a disease-free two-cycle. The first, recently proposed in \cite{peterson2026modeling}, is a stage-structured model describing the transmission of \textit{Borrelia burgdorferi}, the causative agent of Lyme disease, between \textit{Ixodes scapularis} ticks and multiple vertebrate hosts. Infection is transmitted between ticks and hosts through tick feeding, with infected ticks subsequently capable of transmitting the pathogen to hosts during later blood meals. The second is a generic host-parasitoid model adapted from \cite{beddington1975dynamic}, in which hosts transmit infection through horizontal and imperfect vertical transmission, while parasitoids may acquire infection horizontally from infected hosts or vertically from infected parasitoids. This model could represent a system of Cynipid gall wasps and their parasitoids, with a pathogen such as a \textit{Wolbachia} circulating within the host and parasitoid populations \citep{zhao2021wolbachia, hou2020wolbachia, huigens2004natural}. This system provides a setting in which to investigate the relationship between demographic bistability and pathogen invasion without the added machinery of a complete host--vector--host transmission pathway. Although the two models differ in their biological motivation and transmission mechanisms, both exhibit the same qualitative demographic structure: a stable disease-free fixed point with stable disease-free cycles. 

By comparing reproduction numbers evaluated at the fixed point and cyclic disease-free attractors, we investigate how the underlying demographic dynamics and transmission structure influence pathogen invasion. In particular, we ask whether a single threshold quantity adequately characterizes invasion when multiple disease-free attractors of different types coexist, and under what conditions the resulting invasion thresholds differ. 

\section{Tick-Host Model}

\subsection{Model Development}
\label{development}
In this section, we summarize the first epidemiological model analyzed in this paper, which is a tick-host system modeling \textit{B. burgdorferi} transmission in the northeastern U.S. The model combines the demographic framework developed in \cite{peterson2026modeling} with the infection dynamics introduced in \cite{peterson2026role}. The model includes six host groups:
\begin{itemize}
    \item Group 1: White-footed mice,
    \item Group 2: Chipmunks,
    \item Group 3: Short-tailed shrews, \textit{Sorex} shrews, and red-backed voles
    \item Group 4: Squirrels,
    \item Group 5: Five-lined skinks and eastern fence lizards
    \item Group 6: Deer, opossums, and raccoons,
\end{itemize}
and three tick life stages: larvae ($L$), nymphs ($N$), and adults ($A$).

\subsubsection{Modeling Assumptions}\label{assumptions}
Below we summarize key assumptions that are most relevant to the work presented here; a complete list is given in the studies this work is based on \citep{peterson2026modeling, peterson2026role}. 
\begin{itemize}
    \item Tick life stages feed sequentially throughout the year, with nymphs feeding in spring, larvae in summer, and adults in fall \citep{eisen2025seasonal}.
    \item Larvae and nymphs feed on all host groups, whereas adults feed only on the medium- and large-bodied mammal species included in host group 6 \citep{CDC2022}.
    \item Ticks' host-finding success decreases with the abundance of the previous tick cohort, representing delayed host responses, such as nest relocation, habitat avoidance, or the development of resistance following repeated infestations \citep{gibson2022evolution, stanback2001within, klemme2016learned}.
    \item Hosts in group 6 are treated as incompetent reservoirs and therefore do not participate in pathogen transmission \citep{logiudice2003ecology}.
    \item Infection is transmitted only through horizontal transmission during tick blood meals \citep{rollend2013transovarial, kurokawa2020interactions, magnarelli1987transovarial,voordouw2015co, piesman2001efficacy}.
    \item Infection persists throughout the lifetime of both ticks and hosts, with no increased death due to infection \citep{voordouw2015lyme, barbour1993biological, ostfeld2000biodiversity}.
\end{itemize}

\subsubsection{The Mathematical Model}\label{model}

Under the assumptions noted in Section~\ref{assumptions} and in \cite{peterson2026modeling}, the resulting model is

\begin{equation}\label{fullsystem}
\begin{aligned}
    L_S\left(t+1\right)&=\frac{r\frac{A\left(t\right)}{2}}{\frac{\frac{A\left(t\right)}{2}}{H_6\left(t\right)}+K}s_LS_L(t)\left(1-\sum_{i=1}^5 p_{i,L}(t) \beta_{i,L} \frac{H_{i_I}\left(t\right)}{H_i\left(t\right)+\frac{\Lambda_i}{k_i}}\right)\\
    L_I\left(t+1\right)&=\frac{r\frac{A\left(t\right)}{2}}{\frac{\frac{A\left(t\right)}{2}}{H_6\left(t\right)}+K}s_LS_L(t)\left(\sum_{i=1}^5 p_{i,L}(t) \beta_{i,L} \frac{H_{i_I}\left(t\right)}{H_i\left(t\right)+\frac{\Lambda_i}{k_i}}\right)\\
    N_S\left(t+1\right)&=s_NS_N(t)L_S\left(t\right)\left(1-\sum_{i=1}^5p_{i,N}(t)\beta_{i,N}\frac{H_{i_I}\left(t\right)e^{-\mu_i}}{H_i\left(t\right)}\right)\\
    N_I\left(t+1\right)&=s_NS_N(t)\left(L_I\left(t\right)+L_S\left(t\right)\left(\sum_{i=1}^5p_{i,N}(t)\beta_{i,N}\frac{H_{i_I}\left(t\right)e^{-\mu_i}}{H_i\left(t\right)}\right)\right)\\
    A_S\left(t+1\right)&=s_AS_A(t)s_NS_N(t)L_S\left(t\right)\left(1-\sum_{i=1}^5p_{i,N}(t)\beta_{i,N}\frac{H_{i_I}\left(t\right)e^{-\mu_i}}{H_i\left(t\right)}\right)\\
    A_I\left(t+1\right)&=s_AS_A(t)s_NS_N(t)\left(L_I\left(t\right)+L_S\left(t\right)\left(\sum_{i=1}^5p_{i,N}(t)\beta_{i,N}\frac{H_{i_I}\left(t\right)e^{-\mu_i}}{H_i\left(t\right)}\right)\right)\\~\\
    H_{i_S}\left(t+1\right)&=\left(\left(H_{i_S}(t)+\frac{\Lambda_i}{2}\left(1+e^{\frac{\mu_i}{2}}\right)\right)e^{-\mu_i}\right)e^{-\frac{p_{i,N}(t)\alpha_{i}s_NS_N(t)L_I\left(t\right)}{H_i\left(t\right)}}\quad \text{for $i=1,2,3,4$}\\
    H_{i_I}\left(t+1\right)&=\left(H_{i_I}\left(t\right)+\left(H_{i_S}(t)+\frac{\Lambda_i}{2}\left(1+e^{\frac{\mu_i}{2}}\right)\right)\left(1-e^{-\frac{p_{i,N}(t)\alpha_{i}s_NS_N(t)L_I\left(t\right)}{H_i\left(t\right)}}\right)\right)e^{-\mu_i} \quad \text{for $i=1,2,3,4$}\\
    H_{5_S}\left(t+1\right)&=\left(\left(H_{5_S}\left(t\right)+\Lambda_5\right)e^{-\mu_5}\right)e^{-\frac{p_{5,N}(t)\alpha_{5}s_NS_N(t)L_I\left(t\right)}{H_5\left(t\right)}}\\
    H_{5_I}\left(t+1\right)&=\left(H_{5_I}\left(t\right)+\left(H_{5_S}\left(t\right)+\Lambda_5\right)\left(1-e^{-\frac{p_{i,N}\left(t\right)\alpha_is_NS_N\left(t\right)L_I\left(t\right)}{H_5\left(t\right)}}\right)\right)e^{-\mu_5}\\
    H_6\left(t+1\right)&=\left(H_6(t)e^{-\frac{\mu_6}{2}}+\Lambda_6\right)e^{-\frac{\mu_6}{2}}
\end{aligned}
\end{equation}
where 
\begin{align*}
S_L(t) &= 1 - e^{- \frac{z_L}{L(t)}\left(\sum_{i=1}^5 b_{i,L} \left(H_i(t) + \frac{\Lambda_i}{k_i}\right) + b_{6,L} \left(H_6(t)e^{- \mu_6/2} + \Lambda_6\right)\right)}, \\
S_N(t) &= 1 - e^{- \frac{z_N}{N(t)}\left(\sum_{i=1}^5 b_{i,N}\sigma_{i,N} H_i(t) + b_{6,N}\sigma_{6,N} H_6(t)e^{- \mu_6/2}\right)}, \\
S_A(t) &= 1 - e^{- \frac{z_A}{A(t)}b_{6,A}\sigma_{6,A} H_6(t)}, \\
p_{i,L}(t) &= \frac{b_{i,L} \left(H_i(t) + \frac{\Lambda_i}{k_i}\right) e^{- \mu_i/2}}{\sum_{j=1}^5 b_{j,L} \left(H_j(t) + \frac{\Lambda_j}{k_j}\right) e^{- \mu_j/2} + b_{6,L} H_6(t)}, \quad i = 1,\dots,5, \\
p_{6,L}(t) &= \frac{b_{6,L} H_6(t)}{\sum_{j=1}^5 b_{j,L} \left(H_j(t) + \frac{\Lambda_j}{k_j}\right) e^{- \mu_j/2} + b_{6,L} H_6(t)}, \\
p_{i,N}(t) &= \frac{b_{i,N}\sigma_{i,N} H_i(t)}{\sum_{j=1}^5 b_{j,N}\sigma_{j,N} H_j(t) + b_{6,N}\sigma_{6,N} H_6(t)e^{- \mu_6/2}}, \quad i = 1,\dots,5, \\
p_{6,N}(t) &= \frac{b_{6,N}\sigma_{6,N} H_6(t)e^{- \mu_6/2}}{\sum_{j=1}^5 b_{j,N}\sigma_{j,N} H_j(t) + b_{6,N}\sigma_{6,N} H_6(t)e^{- \mu_6/2}}, \\
k_i &= 2 \quad \text{for } i = 1,2,3,4; \quad k_5 = 1. 
\end{align*}

The full yearly model is derived by sequencing ordered, nonoverlapping biological events and then composing the corresponding events. The sequence of events is given in Appendix~\ref{ap:timeline} and Table~\ref{timeline}. A complete explanation of the model derivation and structure is given in \cite{peterson2026modeling} and \cite{peterson2026role}. Table~\ref{param_table} lists the model parameters and state variables with descriptions and units. 

\begin{table}[h!]
    \centering
    \setlength{\tabcolsep}{5.5pt}
    \begin{tabular}{cp{7.5cm}c}
    \toprule
    \textbf{State Var.} & \textbf{Definition} & \textbf{Units}\\
    \midrule
    $L_{S,I}$ & Susceptible and infected larvae & larvae/ha\\
    $N_{S,I}$ & Susceptible and infected nymphs & nymphs/ha\\
    $A_{S,I}$ & Susceptible and infected adults & adults/ha\\
    $H_{1_{S,I}}$ & Susceptible and infected white-footed mice & mice/ha\\
    $H_{2_{S,I}}$ & Susceptible and infected chipmunks & chipmunks/ha\\
    $H_{3_{S,I}}$ & Susceptible and infected short-tailed shrews & shrews/ha\\
    $H_{4_{S,I}}$ & Susceptible and infected squirrels & squirrels/ha\\
    $H_{5_{S,I}}$ & Susceptible and infected lizards& lizards/ha\\
    $H_6$ & Deer & deer/ha\\
    \midrule
    \textbf{Parameter} & \textbf{Definition} & \textbf{Units} \\
    \midrule
    $\Lambda_{i}$  & Birth/recruitment of host group $i$ & host $i$/ha\\
    $\mu_{i}$ & Natural death of host $i$ & dimensionless\\
    $s_j$ & Natural survival of ticks in life stage $j$ & dimensionless\\
    $z_j$ & Host-finding calibration constant & dimensionless\\
    $r$ & Fecundity & larvae/deer\\
    $K$ & Tick-host ratio at which tick reproduction is at half its maximum & adults/deer\\
    $\beta_{i,j}$ & Proportion of ticks in life stage $j$ that become infected after feeding on an infected host from group $i$ & dimensionless\\
    $\alpha_i$ & Proportion of hosts from group $i$ that become infected after a bite from an infected nymph & host $i$/nymph\\
    $b_{i,j}$ & Relative seasonal feeding potential of tick stage $j$ for host group $i$& 1/host $i$\\
    $\sigma_{i,j}$ & Host density scaling constant & dimensionless\\
    \bottomrule
    \end{tabular}
    \caption{State variables and parameters for Model \eqref{fullsystem} with their descriptions and units. $i \in \{1,2,3,4,5,6\}$ and $j \in \{L,N,A\}$. }
    \label{param_table}
\end{table}

\subsection{Baseline Demographic Dynamics}
Before considering pathogen transmission, we first characterize the underlying tick-host dynamics in the absence of infection. Summing the susceptible and infected compartments of each population in Model~\eqref{fullsystem} recovers the demographic system
\begin{equation}\label{tickhostsystem}
\begin{aligned}
    L(t+1)&=\frac{rA(t)/2}{\frac{A(t)/2}{H_6(t)}+K}s_LS_L(t),\\
    N(t+1)&=s_NS_N(t)L(t),\\
    A(t+1)&=s_AS_A(t)N(t+1),\\~\\
    H_{i}(t+1)&=\left(H_{i}(t)+\frac{\Lambda_i}{2}\left(1+e^{\frac{\mu_i}{2}}\right)\right)e^{-\mu_i}, \quad \text{for $i=1,2,3,4$},\\
    H_{5}(t+1)&=\left(H_{5}(t)+\Lambda_5\right)e^{-\mu_5},\\
    H_6\left(t+1\right)&=\left(H_6(t)e^{-\frac{\mu_6}{2}}+\Lambda_6\right)e^{-\frac{\mu_6}{2}},
\end{aligned}
\end{equation}
As shown in \cite{peterson2026modeling}, the persistence of the tick population is governed by the demographic reproductive number \[\mathcal{R}_d=\frac{rS}{K},\] where $S=\frac{1}{2}s_Ls_Ns_A$. This quantity summarizes the effects of adult reproduction and  stage-specific survival and can be interpreted as the expected lifetime reproductive output of an adult female tick \citep{cushing1994net, cushing1989ebenman}. When $\mathcal{R}_d<1$, insufficient reproduction and survival prevent population persistence, and the system approaches the extinction fixed point. 

The dynamics at the threshold $\mathcal{R}_d=1$ are characterized by two simultaneous bifurcations. At this point, the extinction fixed point undergoes a transcritical bifurcation associated with an eigenvalue of $+1$, giving rise to a stable positive fixed point. At the same parameter value, an eigenvalue of $-1$ produces a period-doubling bifurcation and an unstable synchronous 2-cycle. Along this cycle, the tick population alternates between a larval-only state ($N=A=0$) and a state containing only nymphs and adults ($L=0$). This produces a single-cohort (synchronous) cycle, reflecting the two-year structure of the \textit{I. scapularis} life cycle. 

For $\mathcal{R}_d>1$, dynamics become more complicated. For an intermediate range of $\mathcal{R}_d$ values, a stable, strictly positive 2-cycle emerges and collides with the synchronous 2-cycle, handing off stability to the synchronous cycle. Because the fixed point is still stable in this region, the system exhibits bistability between the fixed point and the stable cycle. For sufficiently large values of $\mathcal{R}_d$, the fixed point becomes unstable, and the synchronous 2-cycle is the only stable attractor.

\begin{figure*}[ht!]
\centering
\includegraphics[width=\linewidth]{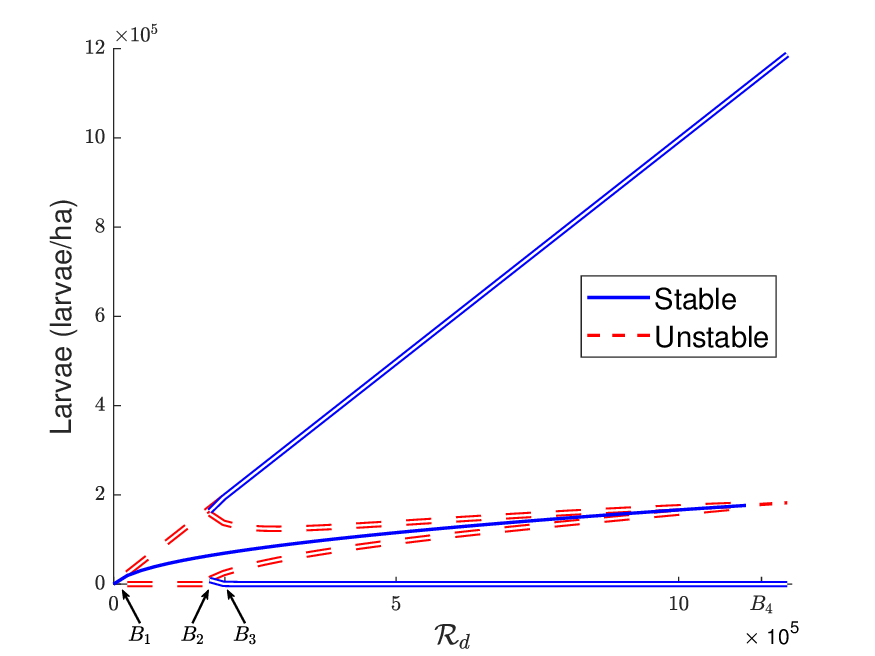}
    \caption{Larval tick population dynamics with respect to the demographic reproductive number $\mathcal{R}_d$. The bifurcation labels along the $x$-axis are described in the main text.}
    \label{pop_dynamics}
\end{figure*}

These dynamics are shown in Figure~\ref{pop_dynamics} for the larval tick population. The dynamics of the nymph and adult populations are qualitatively similar. The bifurcations described above are labeled along the $x$-axis. $B_1$ corresponds to the $+1/-1$ bifurcation at $\mathcal{R}_d=1$, $B_2$ denotes the fold bifurcation of 2-cycles that creates a stable, strictly positive 2-cycle, $B_3$ represents the transcritical bifurcation of 2-cycles, and $B_4$ denotes the subcritical flip bifurcation that destabilizes the positive existence fixed point. The bifurcation at $B_1$ is not displayed in the figure because the scale of the horizontal axis makes it visually indistinguishable from zero.

Because infection dynamics occur on top of these demographic states, the disease system should be analyzed under each stable attractor. In particular, when tick populations oscillate in a stable 2-cycle, transmission becomes periodic, and important disease metrics, such as the basic reproductive number, must be computed over one full demographic period (i.e., two one-year time steps). The following sections therefore consider infection dynamics under both convergence to a fixed point and convergence to a cycle. 

\subsection{Transmission Dynamics}\label{transmissiondynamics}
In this section, we analyze the infection dynamics of Model~\eqref{fullsystem} under both equilibrium (Section~\ref{BRN_FP}) and oscillatory (Section~\ref{BRN_2}) dynamics. We derive expressions for the basic reproductive number, $\mathcal{R}_0$, under each of these conditions.

\subsubsection{Basic Reproductive Number - Constant Populations}\label{BRN_FP}
To determine the basic reproductive number under the assumption of convergence to a stable fixed point (denoted $\mathcal{R}_{0_1}$), we analyze the infected populations of Model \eqref{fullsystem} after all demographic variables have reached equilibrium:
\begin{equation}\label{BRNsystem}
\begin{aligned}
    L_I\left(t+1\right)&=\frac{r\frac{A_{\infty}}{2}}{\frac{\frac{A_{\infty}}{2}}{H_{6_{\infty}}}+K}s_LS_{L_{\infty}}\left(\sum_{i=1}^5 p_{{i,L}_{\infty}} \beta_{i,L} \frac{H_{i_I}\left(t\right)}{H_{i_{\infty}}+\frac{\Lambda_i}{k_i}}\right)\\
    N_I\left(t+1\right)&=s_NS_{N_\infty}\left(L_I\left(t\right)+L_S\left(t\right)\left(\sum_{i=1}^5p_{{i,N}_\infty}\beta_{i,N}\frac{H_{i_I}\left(t\right)e^{-\mu_i}}{H_{i_\infty}}\right)\right)\\
    A_I\left(t+1\right)&=s_AS_{A_\infty}s_NS_{N_\infty}\left(L_I\left(t\right)+L_S\left(t\right)\left(\sum_{i=1}^5p_{{i,N}_\infty}\beta_{i,N}\frac{H_{i_I}\left(t\right)e^{-\mu_i}}{H_{i_\infty}}\right)\right)\\
    H_{i_I}\left(t+1\right)&=\left(H_{i_I}\left(t\right)+ \left(H_{i_S}(t)+\lambda_i\right)\left(1-e^{-\frac{p_{{i,N}_\infty}\alpha_{i}s_NS_{N_\infty}L_I\left(t\right)}{H_{i_\infty}}}\right)\right)e^{-\mu_i},
\end{aligned}
\end{equation}
where
\begin{align*}
    \lambda_i = \frac{\Lambda_i}{2}\left(1+e^{\frac{\mu_i}{2}}\right) \quad \text{for $i \in \{1,2,3,4\}$}, \quad
    \lambda_5 = \Lambda_5.
\end{align*}
At the demographic fixed point, the feeding-success probabilities and host contact distributions no longer vary with time. We therefore write these quantities as $S_{j_\infty}$ and $p_{{i,j}_{\infty}}$, respectively.

Since the infected larval and host equations do not depend on nymphs or adults, these equations decouple, and the basic reproductive number is calculated using the larvae-host subsystem. This simplification arises from the modeling assumption that adults feed exclusively on host group 6, which is assumed to be incompetent for \textit{B. burgdorferi} infection, meaning no new infections (that are tracked in our model) are generated from nymphs or adults.

Using the next-generation operator formulation for discrete-time systems \citep{allen2008basic} (see Appendix~\ref{ap:BRN}), we obtain
\begin{align*}
\mathcal{R}_{0_1} &= \sqrt{
        \sum_{i=1}^5 R_iS_{L_\infty}S_{N_\infty}\frac{e^{-\mu_i}}{1-e^{-\mu_i}}},
\end{align*}
where 
\begin{align*}
    R_i&=\frac{r\frac{A_\infty}{2}}{\frac{\frac{A_\infty}{2}}{H_{6_\infty}}+K}s_Ls_N\frac{p_{i,L}p_{i,N}\alpha_i\beta_{i,L}}{H_{i_\infty}+\frac{\Lambda_i}{2}}\frac{H_{i_\infty}+\lambda_i}{H_{i_\infty}}.
\end{align*}
The product $R_iS_{L_\infty}S_{N_\infty}$ quantifies the transmission pathway associated with host group $i$, accounting for infection transfer through the larva-host-larva sequence over one tick life cycle. Summing this contribution across the competent host groups provides a measure of the overall transmission potential generated by the host community. Host survival enters through the factor $\frac{e^{-\mu_i}}{1-e^{-\mu_i}}$, which accounts for the cumulative contribution of host survival to transmission. 

From the expression above, $\mathcal{R}_{0_1}<1$ if and only if $\mathcal{R}_{0_1}^2<1$, or
\begin{align*}
    \hat{\mathcal{R}}_{0_1}=\sum_{i=1}^5R_iS_{L_\infty}S_{N_\infty}\frac{e^{-\mu_i}}{1-e^{-\mu_i}}<1.
\end{align*}
By the next-generation operator method, the disease-free fixed point (where all infected populations are zero) is locally asymptotically stable whenever this condition holds. Analysis given in Appendix~\ref{ap:ee} shows the existence and stability of an endemic fixed point when $\hat{\mathcal{R}}_{0_1}>1$.

\subsubsection{Basic Reproductive Number - Oscillating Populations}\label{BRN_2}

When populations oscillate rather than converge to a fixed point value, the basic reproductive number (denoted here as $\mathcal{R}_{0_2}$) must be computed over one full period of the demographic cycle instead of a single time step as done in Section~\ref{BRN_FP}.

Several approaches exist for computing the basic reproductive number in periodic systems. One approach, following \cite{caswell2009stage}, \cite{van2019demographic}, and \cite{cushing2012net}, linearizes the infection subsystem at each point of the cycle and considers the product of the corresponding Jacobian matrices over one full period. From this product, transmission and transition matrices $F$ and $T$ can be identified, and the basic reproductive number is defined as the spectral radius 
\[ \mathcal{R}_{0_2}=\rho(F(I-T)^{-1}).\]
Because the Jacobian product depends on the ordering of the cycle points, this procedure can yield multiple values corresponding to the choice of starting point in the cycle. However, cyclic permutations of the product have the same nonzero eigenvalues \citep{bhatia2002eigenvalues}. Thus, all possible Jacobian product matrices yield the same stability threshold, so all the basic reproductive numbers developed from them must yield equivalent thresholds. 

An alternative formulation proposed by \cite{ackleh2022modeling} (though the derivation of this approach is not discussed in their study) defines the basic reproductive number as the spectral radius of the product of next-generation matrices computed at each point in the cycle. That is,
\[\mathcal{R}_{0_2}=\rho(F_2(I-T_2)^{-1}F_1(I-T_1)^{-1}),\]
where $F_1$ and $T_1$ are taken from the Jacobian evaluated at one point of the 2-cycle, and $F_2$ and $T_2$
are taken from the Jacobian evaluated at the second point.

In our case, this formulation yields the same value as the Jacobian-product method noted earlier for one of the two possible matrix multiplication orders. Again, because eigenvalues are preserved under commutation of products of square matrices, this method yields the same spectral radius regardless of the order of the next-generation matrices being multiplied here, i.e., independently of the choice of starting point.

In either case, we linearize the infection subsystem at both points of the 2-cycle and consider the product of the corresponding Jacobian matrices to determine $\mathcal{R}_{0_2}$. Because a closed-form expression for the strictly positive 2-cycle cannot be obtained, the analysis presented here focuses on the synchronous case. The same procedure, however, can be applied numerically to the positive cycle, and the dynamics are shown quantitatively in Section~\ref{quantitative_analysis}. For the synchronous 2-cycle, the two points are
\begin{align*}
    (L_1,N_1,A_1) &= \left(\frac{2H_{6_\infty}(rS-K)}{s_Ns_A},0,0\right),\\
    (L_2,N_2,A_2) &= \left(0,\frac{2H_{6_\infty}(rS-K)}{s_A},2H_{6_\infty}(rS-K)\right).
\end{align*}
Although we number these points for convenience, the ordering of the two states depends on the initial conditions. Therefore, we compute the basic reproductive number under both possible orderings of the synchronous 2-cycle. These scenarios correspond to (1) initializing the cycle with larvae present and no nymphs or adults, in which case
\begin{align*}
    \mathcal{R}_{0_2} &= \frac{1}{2} \left(\sum_{i=1}^5R_ie^{-\mu_i} + \sqrt{
        \left(\sum_{i=1}^5R_ie^{-\mu_i}\right)^2 + 4 \sum_{i=1}^5 R_i\frac{e^{-3\mu_i}}{1-e^{-2\mu_i}}}\right),
\end{align*}
and (2) initializing the cycle with no larvae but with nymphs and adults present, in which case
\begin{align*}
    \hat{\mathcal{R}}_{0_2}= \sum_{i=1}^5 R_i\frac{e^{-\mu_i}}{1-e^{-2\mu_i}}.
\end{align*}
The quantity $R_i$ retains the same definition and biological interpretation as in the constant population case, representing the contribution of host group $i$ to larva-host-larva transmission. The term $\frac{e^{-\mu_i}}{1-e^{-2\mu_i}}$ that appears here is the host survival factor over successive full cycles and therefore captures repeated transmission opportunities that occur every other year under oscillatory population dynamics. 

The difference in form between the expressions for $\mathcal{R}_{0_2}$ and $\hat{\mathcal{R}}_{0_2}$ arises from the structure of the infection pathways at the start of the demographic cycle. When the 2-cycle is initialized with $L>0$, larvae are present in the first phase of the cycle. Although larvae are not born infected (or we assume the proportion that are is negligible \citep{rollend2013transovarial, magnarelli1987transovarial}), they can immediately feed on infected hosts and potentially acquire the pathogen. This creates infected nymphs in the following time step, which can then transmit infection back to hosts. As a result, transmission occurs through a two-step feedback loop: infected hosts infect larvae, those larvae develop into infected nymphs, and the infected nymphs subsequently infect new hosts. The expression for $\mathcal{R}_{0_2}$ therefore reflects the full host-tick-host transmission pathway and appears as the larger square root expression.

In contrast, when the cycle is initialized with $L=0$, no larvae are present in the first phase of the cycle. Because the larval stage is absent, infected hosts initially have no larvae to infect. The only transmission that can occur in that phase (that we track in our model) is infection of hosts by the currently feeding nymphs. The resulting infected hosts must then persist to the next phase, when larvae appear and can feed on them. Because the larval infection pathway is effectively delayed by one time step, the resulting expression for $\hat{\mathcal{R}}_{0_2}$ simplifies to the sum term.

Although these expressions are different algebraically, the threshold conditions obtained from $\mathcal{R}_{0_2}<1$ and $\hat{\mathcal{R}}_{0_2}<1$ are indeed identical:
\begin{align*}
    \sum_{i=1}^5R_i\frac{1}{1+e^{-\mu_i}}\frac{e^{-\mu_i}}{1-e^{-\mu_i}}<1.
\end{align*}
This condition ensures that infection cannot replace itself over one full demographic cycle. Thus, while the two expressions give different values numerically, they are equivalent for thresholding purposes. This result agrees with the biological intuition that long-term persistence depends on whether the infection can replace itself over the full two-year cycle, regardless of where in the cycle it begins.

When $\mathcal{R}_{0_2}<1$, the disease-free synchronous 2-cycle is locally asymptotically stable by the periodic next-generation operator method. 

Comparing and summarizing the basic reproductive numbers under constant and oscillatory dynamics, we have
\begin{align*}
    \mathcal{R}_{0_1} &= \sqrt{
        \sum_{i=1}^5 R_iS_{L_\infty}S_{N_\infty}\frac{e^{-\mu_i}}{1-e^{-\mu_i}}},\\
    \mathcal{R}_{0_2} &= \frac{1}{2} \left(\sum_{i=1}^5R_ie^{-\mu_i} + \sqrt{
        \left(\sum_{i=1}^5R_ie^{-\mu_i}\right)^2 + 4 \sum_{i=1}^5 R_ie^{-\mu_i}\frac{e^{-2\mu_i}}{1-e^{-2\mu_i}}}\right),\\ 
        \hat{\mathcal{R}}_{0_1}&= \sum_{i=1}^5R_iS_{L_\infty}S_{N_\infty}\frac{e^{-\mu_i}}{1-e^{-\mu_i}},\\
    \hat{\mathcal{R}}_{0_2}&= \sum_{i=1}^5R_i\frac{1}{1+e^{-\mu_i}}\frac{e^{-\mu_i}}{1-e^{-\mu_i}},
\end{align*}
with the corresponding threshold conditions
\begin{align*}
\mathcal{R}_{0_1}<1 &\iff \hat{\mathcal{R}}_{0_1}=\sum_{i=1}^5R_iS_{L_\infty}S_{N_\infty}\frac{e^{-\mu_i}}{1-e^{-\mu_i}}<1,\\
    \mathcal{R}_{0_2}<1 &\iff \hat{\mathcal{R}}_{0_2}=\sum_{i=1}^5R_i\frac{1}{1+e^{-\mu_i}}\frac{e^{-\mu_i}}{1-e^{-\mu_i}}<1.
\end{align*}
Note that both conditions share the common factor 
\begin{align*}
    R_i\frac{e^{-\mu_i}}{1-e^{-\mu_i}},
\end{align*}
which represents the expected number of secondary infections generated through repeated host survival and transmission opportunities. The difference between the two cases arises from the additional weighting terms. In the fixed point case, transmission is constrained by the factors $S_{L_\infty}$ and $S_{N_\infty}$, which represent the efficiency of larval and nymphal feeding success. Because these proportions are less than one, they reduce the expected number of successful host-tick contacts and therefore decrease the overall transmission potential.

In contrast, these feeding success terms do not appear explicitly in the synchronous 2-cycle case because, when the system is evaluated at the two points of the cycle, either the larval or the nymphal population is zero. In those phases the corresponding feeding stage is absent, so $S_{L_\infty}$ and $S_{N_\infty}$ evaluate to one and do not constrain transmission in the linearized infection subsystem. As a result, the dominant weighting in the synchronous 2-cycle case instead arises from host mortality through the factor $\frac{1}{1+e^{-\mu_i}}$, which captures the proportion of hosts that were alive the previous year.

These results show that demographic oscillations modify the weighting of host survival and feeding success in the basic reproductive number. 

\subsection{Influence of Demographic Dynamics on Infection}\label{quantitative_analysis}

In this section, we investigate how different demographic dynamics influence pathogen transmission through their effects on the basic reproductive number. In particular, we compare values computed at the positive existence fixed point, $\mathcal{R}_{0_1}$, with those obtained when the demographic dynamics converge to a periodic 2-cycle, $\mathcal{R}_{0_2}$ (see Sections~\ref{BRN_FP} and~\ref{BRN_2} for details). Since the basic reproductive number is defined only for stable demographic dynamics, these quantities are computed only when the corresponding dynamics are stable.

All numerical work presented here was conducted in MATLAB\textsuperscript{\textregistered} using \verb|fsolve| and the parameter values listed in Table~\ref{paramvaluesnorth}. An in-depth discussion of these parameters can be found in \cite{peterson2026host, peterson2026modeling, peterson2026role}. These values were selected based on a literature review and represent, to the extent possible, biologically realistic values for tick populations in the northeastern US. 

The larval and nymphal host-finding success calibration constants, $z_L$ and $z_N$, were altered for the purposes of this work. These parameters represent the responsiveness of larvae and nymphs to the host-to-tick ratio from the previous year and have not been measured experimentally. In our previous work, we calibrated $z_L$ and $z_N$ using observed host and tick densities together with estimates of overall tick survival \cite{peterson2026modeling}. For the present study, which aims to illustrate the potential consequences of having multiple reproductive numbers, we scaled these calibrated values by a factor of $0.15$. Thus, the values reported in Table~\ref{paramvaluesnorth} are $0.15$ times the originally calibrated values. This scaling was chosen to obtain a parameter set where $\mathcal{R}_{0_1}$ and $\mathcal{R}_{0_2}$ are on opposite sides of the threshold $\mathcal{R}_0=1$ for some values of the demographic reproductive number. While using the original calibrated values of $z_L$ and $z_N$ resulted in both reproductive numbers above one over the range considered, the general trend of $\mathcal{R}_{0_2}$ being larger than $\mathcal{R}_{0_1}$ was still observed. 

\begin{table}[htbp!]
\centering
\resizebox{\textwidth}{!}{%
\begin{tabular}{llrr}
\toprule
\makecell{\textbf{Baseline}\\\textbf{Densities}} & \textbf{Description} & \textbf{Value (Source)} &  \textbf{Units} \\
\midrule
$L(0)$ & Larvae
    & $115{,}000$ \citep{daniels2000estimating}
    & larvae/ha \\
$N(0)$ & Nymphs
    & $11{,}500$ \citep{daniels2000estimating}
    & nymph/ha \\
$A(0)$ & Adults
    & $3{,}450$ \citep{daniels2000estimating}
    & adult/ha \\
$H_{1}(0)$ & Mice
    & $40$ \citep{peterson2026host}
    & mice/ha \\
$H_{2}(0)$ & Chipmunks
    & $30$ \citep{peterson2026host}
    & chipmunks/ha \\
$H_{3}(0)$ & Shrews
    & $59.2$ \citep{peterson2026host}
    & shrews/ha \\
$H_{4}(0)$ & Squirrels
    & $6$ \citep{peterson2026host}
    & squirrels/ha \\
$H_{5}(0)$ & Lizards
    & $27$ \citep{peterson2026host}
    & lizard/ha \\
$H_{6}(0)$ & Deer
    & $0.46$ \citep{peterson2026host}
    & deer/ha \\
\midrule
\textbf{Parameters} & \textbf{Description} & \textbf{Units} \\
\midrule
$\Lambda_1$ & Recruitment of mice
    & $300^{a}$
    & mice/ha \\
$\Lambda_2$ & Recruitment of chipmunks
    & $50$ \citep{snyder1982tamias}
    & chipmunks/ha \\
$\Lambda_3$ & Recruitment of shrews
    & $659.8^{b}$
    & shrews/ha \\
$\Lambda_4$ & Recruitment of squirrels
    & $27^{c}$
    & squirrels/ha \\
$\Lambda_5$ & Recruitment of lizards
    & $36.1^{d}$
    & lizards/ha \\
$\Lambda_6$ & Recruitment of deer
    & $0.32^{e}$
    & deer/ha \\
$s_L$ & Natural survival of larvae
    & $0.43$ \citep{brunner2023off}
    & dimensionless \\
$s_N$ & Natural survival of nymphs
    & $0.55$ \citep{brunner2023off}
    & dimensionless \\
$s_A$ & Natural survival of adults
    & $0.84$ \citep{brunner2023off}
    & dimensionless \\
$\mu_1$ & Natural death of mice
    & $3.12$
    & dimensionless \\
$\mu_2$ & Natural death of chipmunks
    & $1.21$
    & dimensionless \\
$\mu_3$ & Natural death of shrews
    & $3.77$
    & dimensionless \\
$\mu_4$ & Natural death of squirrels
    & $2.36$
    & dimensionless \\
$\mu_5$ & Natural death of lizards
    & $0.85$
    & dimensionless \\
$\mu_6$ & Natural death of deer
    & $0.68$
    & dimensionless \\
$\hat{b}_{1,L}$ & Relative proportion of larvae feeding on mice
    & $404.4$ \citep{ginsberg2021lyme}
    & larvae/mouse \\
$\hat{b}_{2,L}$ & Relative proportion of larvae feeding on chipmunks
    & $245.2$ \citep{ginsberg2021lyme}
    & larvae/chipmunk \\
$\hat{b}_{3,L}$ & Relative proportion of larvae feeding on shrews
    & $483.7$ \citep{ginsberg2021lyme}
    & larvae/shrew \\
$\hat{b}_{4,L}$ & Relative proportion of larvae feeding on squirrels
    & $911.3^{f}$ 
    & larvae/squirrel \\
$\hat{b}_{5,L}$ & Relative proportion of larvae feeding on lizards
    & $35.3$ \citep{ginsberg2021lyme}
    & larvae/lizard \\
$\hat{b}_{6,L}$ & Relative proportion of larvae feeding on deer
    & $1{,}0817.1$ \citep{piesman1979role}
    & larvae/host 6 \\
$\hat{b}_{1,N}$ & Relative proportion of nymphs feeding on mice
    & $20.4$ \citep{ginsberg2021lyme}
    & nymphs/mouse \\
$\hat{b}_{2,N}$ & Relative proportion of nymphs feeding on chipmunks
    & $188.4$ \citep{ginsberg2021lyme}
    & nymphs/chipmunk \\
$\hat{b}_{3,N}$ & Relative proportion of nymphs feeding on shrews
    & $1.9$ \citep{ginsberg2021lyme}
    & nymphs/shrew \\
$\hat{b}_{4,N}$ & Relative proportion of nymphs feeding on squirrels
    & $1{,}527.8^{f}$
    & nymphs/squirrel \\
$\hat{b}_{5,N}$ & Relative proportion of nymphs feeding on lizards
    & $0.4$ \citep{ginsberg2021lyme}
    & nymphs/lizard \\
$\hat{b}_{6,N}$ & Relative proportion of nymphs feeding on deer
    & $4{,}132.2$ \citep{piesman1979role}
    & nymphs/deer \\
$b_{6,A}$ & Relative proportion of adults feeding on deer
    & $1{,}037.5$ \citep{piesman1979role}
    & adults/deer \\
\makecell[l]{$\sigma_{3,N}$\\\vspace{0.1cm}}
& \makecell[l]{Scaling constant for preference of nymphs\\
(versus larval) for host group 3}
& \makecell[r]{$3.59$\\\vspace{0.1cm}}
& \makecell[r]{dimensionless\\\vspace{0.1cm}} \\

\makecell[l]{$\sigma_{5,N}$\\\vspace{0.1cm}}
& \makecell[l]{Scaling constant for preference of nymphs\\
(versus larval) for host group 5}
& \makecell[r]{$1.07$\\\vspace{0.1cm}}
& \makecell[r]{dimensionless\\\vspace{0.1cm}} \\

\makecell[l]{$\sigma_{6,N}$\\\vspace{0.1cm}}
& \makecell[l]{Scaling constant for preference of nymphs\\
(versus larval) for host group 6}
& \makecell[r]{$1.25$\\\vspace{0.1cm}}
& \makecell[r]{dimensionless\\\vspace{0.1cm}} \\

\makecell[l]{$\sigma_{6,A}$\\\vspace{0.1cm}}
& \makecell[l]{Scaling constant for preference of adults\\
(versus larval) for host group 6}
& \makecell[r]{$1.08$\\\vspace{0.1cm}}
& \makecell[r]{dimensionless\\\vspace{0.1cm}} \\
$r$ & Fecundity of adult ticks
    & $2{,}520{,}000^{g}$
    & larvae/host 6 \\
\makecell[l]{$K$\\\vspace{0.1cm}}
& \makecell[l]{Tick-to-host ratio at which tick reproduction is at\\
half its maximum}
& \makecell[r]{$5$\\\vspace{0.1cm}}
& \makecell[r]{adult/host 6\\\vspace{0.1cm}} \\
$z_L$ & Larval host-finding success calibration constant
    & 
    $0.08$
    & dimensionless \\
$z_N$ & Nymphal host-finding success calibration constant
    & 
    $0.11$
    & dimensionless \\
$z_A$ & Adult host-finding success calibration constant
    & $2.96$
    & dimensionless \\
$c$ & Proportion of questing season spent above the leaf litter
    & $0.18$ \citep{arsnoe2019nymphal}
    & dimensionless \\
$\alpha_1$ & Infection probability from nymphs to mice  & $1$ \citep{sze2012borrelia, brunner2008estimating} & mice/nymph \\
    $\alpha_2$ & Infection probability from nymphs to chipmunks  & $1$ & chipmunks/nymph \\
    $\alpha_3$ & Infection probability from nymphs to shrews  & $1$  & shrews/nymph \\
    $\alpha_4$ & Infection probability from nymphs to squirrels  & $1$ & squirrels/nymph \\
    $\alpha_5$ & Infection probability from nymphs to lizards  & $1$ & lizards/nymph \\
    $\beta_{1,L}$ & Infection probability from mice to larvae & $0.91$ \citep{brunner2008estimating} & dimensionless \\ 
    $\beta_{2,L}$ & Infection probability from chipmunks to larvae & $0.57$ \citep{brunner2008estimating} & dimensionless \\
    $\beta_{3,L}$ & Infection probability from shrews to larvae & $0.56$ \citep{brunner2008estimating, dougherty2024role} & dimensionless \\
    $\beta_{4,L}$ & Infection probability from squirrels to larvae & $0.06$ \citep{brunner2008estimating} & dimensionless \\
    $\beta_{5,L}$ & Infection probability from lizards to larvae & $0.02$ \citep{levin1996reservoir} & dimensionless \\
    $\beta_{1,N}$ & Infection probability from mice to nymphs & $0.91$ \citep{brunner2008estimating} & dimensionless \\ 
    $\beta_{2,N}$ & Infection probability from chipmunks to nymphs & $0.57$ \citep{brunner2008estimating} & dimensionless \\
    $\beta_{3,N}$ & Infection probability from shrews to nymphs & $0.56$ \citep{brunner2008estimating, dougherty2024role} & dimensionless \\
    $\beta_{4,N}$ & Infection probability from squirrels to nymphs & $0.06$ \citep{brunner2008estimating} & dimensionless \\
    $\beta_{5,N}$ & Infection probability from lizards to nymphs & $0.02$ \citep{levin1996reservoir} & dimensionless \\
\bottomrule
\end{tabular}
}
\caption{Baseline population densities and parameters estimated from literature for the northeast for Model~\eqref{fullsystem} with their descriptions, estimated values, and units. Parameters without cited sources were calculated as described in \cite{peterson2026modeling}, except for the larval and nymphal host-finding success calibration constants, which were scaled by a factor of $0.15$ for the purposes of this study (see Section~\ref{quantitative_analysis} for details).\\
$^{a}$\citep{morris1986proximate, mehlhop1978population, chitty1941territorial}\\
$^{b}$\citep{innes1990high, whitaker1998mammals, merritt1981clethrionomys, innes1984life}\\
$^{c}$\citep{layne1952genitale, barkalow1970vital, kemp1970dynamics}\\
$^{d}$\citep{COSEWIC2007, seburn1990population, tinkle1972sceloporus}\\
$^{e}$\citep{verme1984physiology, lay1942ecology, johnson2016habitat}\\
$^{f}$\citep{main1982immature}\\
$^{g}$\citep{baker2024feeding, piesman1979role}}
\label{paramvaluesnorth}
\end{table}

\begin{figure*}[ht!]
\centering
    \includegraphics[width=\linewidth]{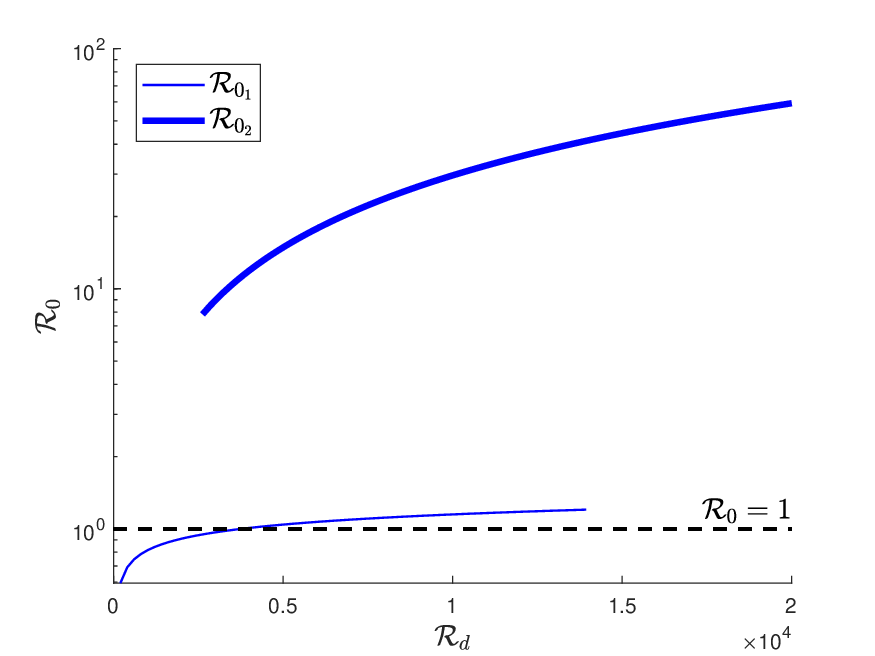}
    \caption{Diagram of $\mathcal{R}_{0_1}$ and $\mathcal{R}_{0_2}$ with respect to the demographic reproductive number $\mathcal{R}_d$ under the parameter set from Table~\ref{paramvaluesnorth}, except for $r$, which was varied to change $\mathcal{R}_d$. In the figure, thin blue lines indicate $\mathcal{R}_{0_1}$ values evaluated at the stable coexistence fixed point, while thick blue lines indicate $\mathcal{R}_{0_2}$ values evaluated at the stable 2-cycle.}
    \label{R0_rSK}
\end{figure*}

Figure~\ref{R0_rSK} illustrates how the basic reproductive numbers vary with the demographic reproductive number $\mathcal{R}_d=\frac{rS}{K}$. In the figure, the thin blue curve represents $\mathcal{R}_{0_1}$, while the thick blue curve represents $\mathcal{R}_{0_2}$. 

For completeness, we also computed the corresponding quantities $\hat{\mathcal{R}}_{0_1}$ and $\hat{\mathcal{R}}_{0_2}$ described in Sections~\ref{BRN_FP} and~\ref{BRN_2}. These exhibit the same qualitative behavior as those shown in Figure~\ref{R0_rSK}, with slightly smaller numerical values. For the 2-cycle case, this difference reflects the role of larvae as the stage that acquires infection from hosts and later transmits it as nymphs. When larvae appear at the beginning of the phase (as in $\mathcal{R}_{0_2}$), hosts can immediately infect larvae, producing infected nymphs in the following step. In contrast, for $\hat{\mathcal{R}}_{0_2}$, the phase begins with no larvae, so transmission does not occur until the second step.

Comparing the two basic reproductive numbers in Figure~\ref{R0_rSK}, the reproductive number computed at the 2-cycle exceeds the fixed-point value across the parameter range considered. Importantly, there is an interval of $\mathcal{R}_d$ values ($\mathcal{R}_d \in (2.6\times10^3,3.7\times10^3)$) where $\mathcal{R}_{0_1}<1$ while $\mathcal{R}_{0_2}>1$ for the same parameters. This indicates that the disease should die out when populations are at their fixed point value but persist when populations oscillate in the 2-cycle.  

\section{Host-Parasitoid Model}

\subsection{Model Development}\label{modeldevelop2}

We now consider the second epidemiological model in this study, which is a modification of a model first proposed by \cite{beddington1975dynamic} and subsequently reformulated and analyzed by \cite{kon2006multiple}. The model extends the classical Nicholson-Bailey host-parasitoid equations \citep{nicholson1935balance} by incorporating density dependence in host reproduction. These models describe the interaction between a host population and its insect parasitoids, where parasitoids require hosts for reproduction. The model is given by
\begin{equation}\label{system2noinfect}
    \begin{aligned}
        E(t+1) &= \lambda E(t)e^{-\mu E(t)}e^{-aL(t)}\\
        L(t+1) &= b\lambda E(t)e^{-\mu E(t)}(1-e^{-aL(t)}).
    \end{aligned}
\end{equation}
Here, the host population is regulated by both intraspecific competition and parasitism, while the parasitoid population depends on the number of hosts successfully parasitized. Our modification is that the density-dependent competition term is also incorporated into the parasitoid equation. We assume that hosts must first survive intraspecific competition for resources before reproducing, and this consequently reduces the number of available host eggs for parasitoid larvae to infest. In the following sections, we extend this model to incorporate infection dynamics and examine how the resulting demographic dynamics influence measures of disease invasion and persistence. 

\subsubsection{Modeling Assumptions}\label{assumptions2}
The following assumptions were used to construct the host-parasitoid-infection model:
\begin{itemize}
    \item Parasitized hosts do not reproduce.
    \item Parasitoids must successfully parasitize a host in order to reproduce.
    \item Hosts and parasitoids have a one-time-step lifespan.
    \item Infection is transmitted horizontally between hosts. 
    \item A proportion of infected hosts transmit infection vertically to their offspring.
    \item Infection in the parasitoid population occurs through imperfect vertical transmission and horizontal transmission from infected hosts. 
    \item Infection does not cause additional mortality in either the host or parasitoid population.
\end{itemize}
Because hosts and parasitoids do not survive from one time step to the next, the infected host and parasitoid populations at time $t+1$ consist entirely of newly recruited infected individuals. Infection in the next generation therefore arises through either vertical transmission from infected parent or horizontal transmission following contact with infected hosts. 

\subsubsection{The Mathematical Model}\label{model2}
Under the assumptions listed in Section~\ref{assumptions2}, we extend Model~\eqref{system2noinfect} to include susceptible and infected classes. The resulting model is
\begin{equation}\label{system2}
    \begin{aligned}
      E_S(t+1) &=\hat{E}_S(t)e^{-aL(t)},\\
      E_I(t+1) &= \hat{E}_I(t)e^{-aL(t)},\\
      L_S(t+1) &= b\left(\hat{E}_S(t)+(1-q)\hat{E}_I(t)\right) \frac{L_S(t)+(1-w)L_I(t)}{L(t)}\left(1-e^{-aL(t)}\right),\\
      L_I(t+1) &= b\left(\left(\hat{E}_S(t)+\hat{E}_I(t)\right)\frac{wL_I(t)}{L(t)}+q\hat{E}_I(t)\frac{L_S(t)+(1-w)L_I(t)}{L(t)}\right)\left(1-e^{-aL(t)}\right),
    \end{aligned}
\end{equation}
where
\begin{align*}
    \hat{E}_S(t) &= \lambda \left(E_S(t)\left(1-\beta\frac{E_I(t)}{E(t)}\right)+(1-v)\left(E_I(t)+E_S(t)\left(\beta\frac{E_I(t)}{E(t)}\right)\right)\right)e^{-\mu E(t)}\\
    \hat{E}_I(t) &= \lambda v\left(\left(E_I(t)+E_S(t)\left(\beta\frac{E_I(t)}{E(t)}\right)\right)\right)e^{-\mu E(t)}
\end{align*}

To construct the final model, similar to the tick--host system, we sequenced the biological event and then composed the corresponding event maps to produce the full model. Appendix~\ref{ap:timeline2} and Table~\ref{timeline2} provide the sequence of events considered here. 

The first two equations describe the recruitment of susceptible and infected hosts, respectively. The factor
\begin{align*}
    \lambda e^{-\mu E(t)}e^{-aL(t)}
\end{align*}
represents the per-capita host recruitment rate after accounting for host reproduction, intraspecific competition, and parasitism. Of the hosts that are infected at the time of reproduction, a proportion $v$ transmit infection vertically to their offspring. Susceptible hosts may also acquire infection through horizontal transmission, represented by the term
\begin{align*}
    \beta \frac{E_I(t)}{E(t)},
\end{align*}
where $\beta$ is the proportion of infectious contacts that result in infection.

The final two equations describe the susceptible and infected parasitoid populations. Parasitoids that do not acquire infection through vertical transmission remain susceptible, represented by the proportion 
\begin{align*}
    \frac{L_S(t)+(1-w)L_I(t)}{L(t)},
\end{align*}
where $w$ is the proportion of infected parasitoids that transmit infection vertically. Susceptible parasitoids are produced when an uninfected host is parasitized or when an infected host is parasitized by a parasitoid that does not acquire infection. This is captured by the term
\begin{align*}
    b(\hat{E}_S(t)+(1-q)\hat{E}_I(t))(1-e^{-aL(t)}),
\end{align*}
where $b$ is  number of parasitoids produced per parasitized host and $\displaystyle (1-e^{-aL(t)})$ is the proportion of hosts that are parasitized.

A parasitoid can acquire infection either through vertical transmission, represented by 
\begin{align*}
    \frac{wL_I(t)}{L(t)},
\end{align*}
or through horizontal transmission when a susceptible parasitoid parasitizes an infected host. The latter pathway is represented by 
\begin{align*}
    q\hat{E}_I(t)\frac{L_S(t)+(1-w)L_I(t)}{L(t)},
\end{align*}
where $q$ is the proportion of parasitoids infesting an infected host egg that become infected.

All parameters and state variables for Model~\eqref{system2} are listed in Table~\ref{param_table2}. 

\begin{table}[h!]
    \centering
    \setlength{\tabcolsep}{5.5pt}
    \begin{tabular}{cp{7.5cm}c}
    \toprule
    \textbf{State Var.} & \textbf{Definition} & \textbf{Units}\\
    \midrule
    $E_{S,I}$ & Susceptible and infected hosts & host\\
    $L_{S,I}$ & Susceptible and infected parasitoids & parasitoids\\
    \midrule
    \textbf{Parameter} & \textbf{Definition} & \textbf{Units} \\
    \midrule
    $\lambda$  & Birth/recruitment of hosts & dimensionless\\
    $\mu$ & Intensity of intraspecific competition in the host population & 1/host\\
    $a$ & Searching efficiency of parasitoids & 1/parasitoid\\
    $b$ & Birth/recruitment of parasitoids & parasitoid/host\\
    $\beta$ & Proportion of hosts that become infected after an infectious host contact & dimensionless\\
    $v$ & Proportion successful vertical transmission in host population & dimensionless\\
    $w$ & Proportion successful vertical transmission in parasitoid population & dimensionless\\
    $q$ & Host-to-parasitoid transmission proportion & dimensionless\\
    \bottomrule
    \end{tabular}
    \caption{State variables and parameters for Model \eqref{system2} with their descriptions and units.}
    \label{param_table2}
\end{table}

\subsection{Baseline Demographic Dynamics}

Similar to the tick-host system, we first examine the demographic behavior of the host-parasitoid system in the absence of infection. Model~\eqref{system2noinfect} has three fixed points:
\begin{itemize}
    \item $E_{00}=(0,0)$
    \item $E_{+0} = \left(\frac{\ln(\lambda)}{\mu},0\right)$
    \item $E_{++}=(H_{\infty},P_{\infty})$.
\end{itemize}
A qualitative stability analysis of these fixed points is provided in Appendix~\ref{ap:fp_stability2}.

Figure~\ref{system2_popdynamics} shows the resulting population dynamics as $\mathcal{R}_d$ varies using parameter values given in Table~\ref{paramvalues2}. These parameters were chosen for illustrative purposes and do not represent a specific biological system. Across the parameter range shown, the system exhibits a sequence of bifurcations and regions of bistability. When $L(t)=0$, Model~\eqref{system2noinfect} reduces to the Ricker map \citep{ricker1954stock}, and thus exhibits the well-known period-doubling behavior (see Appendix~\ref{ap:twocycle2} for details of the existence of the host-only 2-cycle). 

For $\mathcal{R}_d<1$, host reproduction is insufficient to sustain the population, and the extinction fixed point is globally stable. At $\mathcal{R}_d=1$, denoted by $B_1$ in Figure~\ref{system2_popdynamics}, a transcritical bifurcation occurs between the extinction fixed point and the host-only fixed point, resulting in an exchange of stability. As $\mathcal{R}_d$ increases, the host-only fixed point is stable until $\mathcal{R}_d=e^2$, corresponding to $B_2$. At this value, the host-only fixed point loses stability through a period-doubling bifurcation, and a stable host-only 2-cycle emerges.

At $B_3$, a pair of coexistence fixed points emerges through a saddle-node bifurcation. This is followed by a period-doubling bifurcation at $B_4$ where an unstable strictly positive coexistence 2-cycle also emerges, and the coexistence fixed point becomes locally stable. At $B_5$ the unstable branch of the coexistence fixed point exits the state space through the host-only fixed point.

For larger values of $\mathcal{R}_d$, the stable host-only 2-cycle undergoes the expected period-doubling cascade. At $B_6$, the stable 2-cycle loses stability and a stable 4-cycle emerges. A second period-doubling bifurcation occurs at $B_7$, producing a stable 8-cycle. At $B_8$, the stable 8-cycle loses stability, and a stable 16-cycle emerges, although this branch is not shown in Figure~\ref{system2_popdynamics}. For still larger values of $\mathcal{R}_d$, the period-doubling cascade continues.

The parameter region considered here is of particular interest because the demographic dynamics admit multiple attractors, including a stable fixed point and stable cycles. Consequently, the long-term demographic state of the population depends on the initial conditions. This feature provides a useful setting for examining how having multiple disease-free attractors influences the calculation and interpretation of the basic reproductive number.

\begin{figure*}[ht!]
\centering
\begin{subfigure}[t]{0.5\textwidth}
    \centering
    \includegraphics[width=\linewidth]{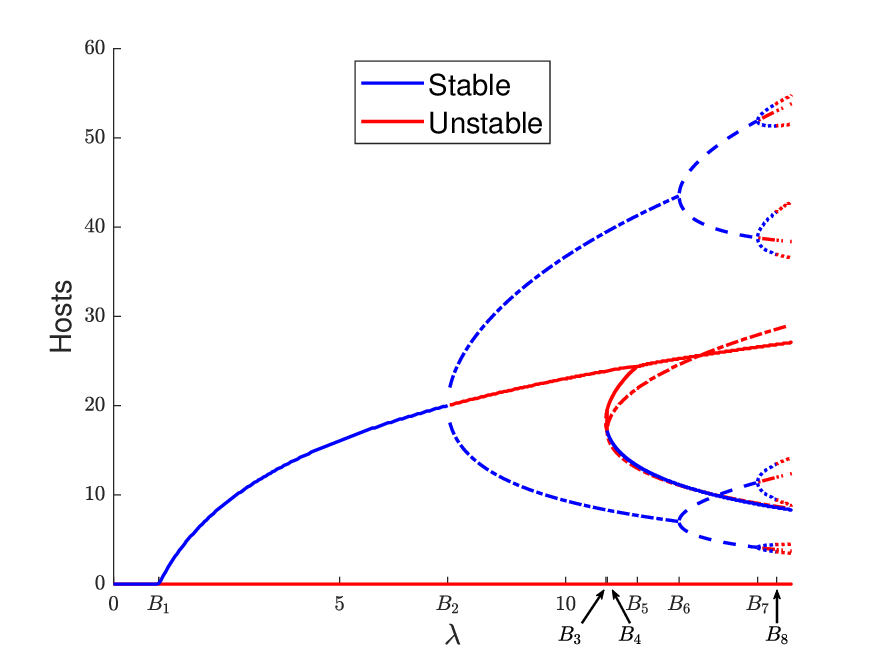}
    \caption{}
    \label{fig:4a}
\end{subfigure}%
\hfill
\begin{subfigure}[t]{0.5\textwidth}
    \centering
    \includegraphics[width=\linewidth]{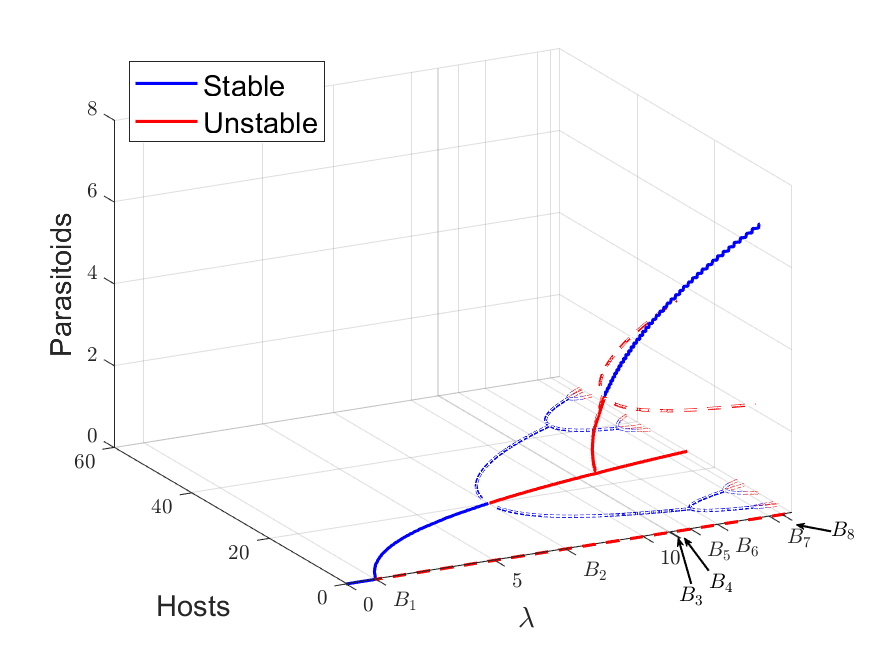}
    \caption{}
    \label{fig:4b}
\end{subfigure}%
\caption{Bifurcation diagrams illustrating the demographic dynamics of Model~\eqref{system2noinfect} as the demographic reproductive number $\mathcal{R}_d=\lambda$ varies using parameters in Table~\ref{paramvalues2}. Subfigure~\ref{fig:4a} shows the host population, while Subfigure~\ref{fig:4b} shows a three-dimensional view of the host and parasitoid populations. In both subfigures, fixed points are shown as solid lines, 2-cycles as a dashed-dotted line, 4-cycles as a dashed line, and 8-cycles as a dotted line. In all cases, lines are blue when stable and red when unstable.}
    \label{system2_popdynamics}
\end{figure*}

\begin{table}[!t]
\centering
\resizebox{\textwidth}{!}{%
\begin{tabular}{llrr}
\toprule
{\textbf{Parameters}} & {\textbf{Description}} & {\textbf{Value}} & \textbf{Units} \\ 
\midrule
    $\lambda$ & Recruitment of hosts & varied & dimensionless \\
    $\mu$ & Intensity of intraspecific competition in the host population  & $0.10$ & 1/host \\
    $a$ & Searching efficiency of parasitoids & $0.29$  & 1/parasitoid \\ 
    $b$ & Recruitment of parasitoids & $0.14$ & parasitoid/host\\
\bottomrule
\end{tabular}
}
\caption{Parameter values used in numerical simulations of Model~\eqref{system2}.}
\label{paramvalues2}
\end{table}

\subsection{Transmission Dynamics}\label{transmissiondynamics2}
\subsubsection{Basic Reproductive Number -- Constant Populations}\label{BRN_FP_2}

We first derive the basic reproductive number for infection introduced into a population converging to a demographic fixed point. Using the next-generation operator method for discrete-time systems \citep{allen2008basic}, we linearize the infected subsystem of Model~\eqref{system2}
    \begin{align*}
      E_I(t+1) &= \hat{E}_I(t)e^{-aL(t)},\\
      L_I(t+1) &= b\left(\left(\hat{E}_S(t)+\hat{E}_I(t)\right)\frac{wL_I(t)}{L(t)}+q\hat{E}_I(t)\frac{L_S(t)+(1-w)L_I(t)}{L(t)}\right)\left(1-e^{-aL(t)}\right),
    \end{align*}
where
\begin{align*}
    \hat{E}_S(t) &= \lambda \left(E_S(t)\left(1-\beta\frac{E_I(t)}{E(t)}\right)+(1-v)\left(E_I(t)+E_S(t)\left(\beta\frac{E_I(t)}{E(t)}\right)\right)\right)e^{-\mu E(t)},\\
    \hat{E}_I(t) &= \lambda v\left(\left(E_I(t)+E_S(t)\left(\beta\frac{E_I(t)}{E(t)}\right)\right)\right)e^{-\mu E(t)},
\end{align*}
about a disease-free fixed point.

In this case, there are two relevant disease-free fixed points: the host-only fixed point ($E_{+0}$) and the coexistence fixed point ($E_{++}$). Because the host-only fixed point is a special case of the coexistence fixed point when $L_\infty=0$, we use the coexistence fixed point in the following derivation.

The $F$ and $T$ matrices in the next-generation operator are given by 
\begin{align*}
    F = 
    \begin{bmatrix}
        (1+\beta)v\lambda e^{-\mu EL_\infty}e^{-aL_\infty} & 0\\
        (1+\beta) v q \lambda b e^{-\mu E_\infty} (1-e^{-a L_\infty}) & \frac{wb\lambda E_\infty e^{-\mu E_\infty}(1-e^{-aL_\infty})}{L_\infty}
    \end{bmatrix},
    \quad
    T = 
    \begin{bmatrix}
        0&0\\
        0&0
    \end{bmatrix}.
\end{align*}
Note that $T$ is the zero matrix since all hosts and parasitoids die after one time step, and thus we have no transition terms. At the fixed point, we can simplify the $F$ matrix to
\begin{align*}
    F = 
    \begin{bmatrix}
        (1+\beta)v\lambda e^{-\mu EL_\infty}e^{-aL_\infty} & 0\\
        (1+\beta) v q \lambda b e^{-\mu E_\infty} (1-e^{-a L_\infty}) & w
    \end{bmatrix}.
\end{align*}
The basic reproductive number is then
\begin{align*}
    \mathcal{R}_{0_1} &= F(I-T)^{-1}= \max\bigg\{(1+\beta)v\lambda e^{-\mu EL_\infty}e^{-aL_\infty},w\bigg\}.
\end{align*}
 
The maximum in the expression for $\mathcal{R}_{0_1}$ arises from the structure of the infection pathways in the model. Infected hosts can generate new infected hosts through horizontal and vertical transmission, while infected parasitoids can generate new infected parasitoids through vertical transmission. However, there is no transmission pathway from infected parasitoids back to hosts. Thus, infection in the parasitoid population cannot contribute to subsequent host infection, and the infection subsystem is triangular. The two diagonal terms therefore represent independent infection growth pathways, and the dominant pathway determines the basic reproductive number. Specifically, $\displaystyle (1+\beta)v\lambda e^{-\mu EL_\infty}e^{-aL_\infty}$ represents the one-step growth factor for infection in the host population, while $w$ represents the one-step growth factor for infection maintained through vertical transmission in the parasitoid population.

Because $w<1$, infection maintained solely through vertical transmission in the parasitoid population cannot independently produce disease persistence. Consequently, the invasion threshold is determined by the host component of $\mathcal{R}_{0_1}$.

\subsubsection{Basic Reproductive Number -- Oscillating Populations}\label{BRN_2_2}

When populations converge to the host-only 2-cycle rather than a fixed point, we derive a corresponding infection threshold using methods developed for periodic discrete-time systems \citep{van2019demographic, cushing2012net, caswell2009stage}. Let the two points of the host-only 2-cycle be denoted by
\begin{align*}
    (E_1,0) = (E_{1_\infty},0), \quad \text{and} \quad (E_2,0) = (E_{2_\infty},0).
\end{align*}
Because the parasitoid population is zero on this cycle, the infected host equation can be evaluated on its own. The Jacobian in this case is a scalar, and evaluating at each point of the 2-cycle gives
\begin{align*}
    J_1 = e^{-\mu E_{1_\infty}}\lambda (\beta+1)v \quad \text{and} \quad J_2 = e^{-\mu E_{2_\infty}}\lambda (\beta+1)v.
\end{align*}
Because the demographic population alternates between these two states, the linearized infection dynamics over one complete period are obtained by multiplying the two one-step infection Jacobians:
\begin{align*}
    J_2J_1 = e^{-\mu E_{1_\infty}}e^{-\mu E_{2_\infty}}\lambda^2 (\beta+1)^2v^2.
\end{align*}
Since infected hosts do not persist between time steps, there is no contribution from infected individuals surviving into the subsequent time step, and therefore $T=0$. We consequently have $F=J_2J_1$ and thus
\begin{align*}
\mathcal{R}_{0_2} = \lambda^2e^{-\mu E_{1_\infty}}e^{-\mu E_{2_\infty}} (\beta+1)^2v^2.
\end{align*}
Here, $\mathcal{R}_{0_2}$ represents infection growth over one complete period of the 2-cycle, and therefore describes growth over two time steps. More generally, if the host-only population converges to a $k$-cycle with points $E_{1_\infty},\dots,E_{k_\infty}$, the corresponding reproduction number is 
\begin{align*}
    \mathcal{R}_{0_k} = \lambda^k (\beta+1)^kv^ke^{-\mu \sum_{j=1}^k E_{j_\infty}}.
\end{align*}
Thus, $\mathcal{R}_{0_k}$ measures the cumulative infection growth over one complete $k$-cycle. To obtain a direct comparison with the basic reproductive number evaluated at a fixed point, which describes growth over a single time step, we define the one-step equivalent reproductive number by taking the $k$th root:
\begin{align*}
    \hat{\mathcal{R}}_{0_k} = \sqrt[1/k]{\mathcal{R}_{0_k}} = \lambda(\beta+1)ve^{-\frac{\mu}{k}\sum_{j=1}^k E_{j_\infty}}.
\end{align*}

The fixed-point and periodic reproductive numbers can now be compared directly. At a disease-free coexistence fixed point, the host demographic fixed point satisfies 
\begin{align*}
    1=\lambda e^{-\mu E_\infty}e^{-a L_\infty},
\end{align*}
or equivalently,
\begin{align*}
    \ln(\lambda)=\mu E_\infty+aL_\infty.
\end{align*}
Consequently, the host component of $\mathcal{R}_{0_1}$ satisfies 
\begin{align*}
    (1+\beta)v\lambda e^{-\mu E_\infty}e^{-aL_\infty}=(1+\beta)v.
\end{align*}
For a host-only $k$-cycle, the demographic equations satisfy 
\begin{align*}
    \ln(\lambda)=\frac{\mu}{k}\sum_{j=1}^kE_{j_\infty}.
\end{align*}
It follows that the one-step equivalent reproduction number for the $k$-cycle also simplifies to
\begin{align*}
    \hat{\mathcal{R}}_{0_k}=(\beta+1)v.
\end{align*}
Therefore, although the demographic populations at the fixed point and along the $k$-cycle may be qualitatively different, the host-driven infection threshold is identical in the two cases. 

Because $w<1$, the fixed-point reproduction number 
\begin{align*}
    \mathcal{R}_{0_1}=\max\{(\beta+1)v,w\}
\end{align*}
crosses one if and only if $(\beta+1)v$ crosses one, which is the same condition for $\mathcal{R}_{0_k}$ to cross one. 

Therefore, demographic bistability between a fixed point and a cycle does not produce different invasion thresholds in this host--parasitoid model. The reproduction numbers themselves may differ because the fixed-point quantity also accounts for the parasitoid vertical-transmission pathway through the maximum with $w$, but this additional pathway cannot change the threshold since $w<1$. Consequently, the fixed point and cyclic demographic attractors yield the same condition for pathogen invasion. 

Although biologically unrealistic, incorporating horizontal transmission from parasitoids to hosts into Model~\eqref{system2} would result in a nontriangular next-generation matrix, leading to a situation similar to that of the tick-host model presented in Section~\ref{model}, with different invasion thresholds associated with each attractor.

\section{Discussion and Conclusions}
\label{discussion}

The basic reproductive number, $\mathcal{R}_0$, is one of the most important quantities used to predict the potential spread of an infectious disease. In this work, we investigated how the interpretation of $\mathcal{R}_0$ is affected by demographic bistability, focusing on systems in which the disease-free population can approach either a stable fixed point or a stable periodic cycle. We considered two discrete-time host-parasite models with infection. The first was a tick-host model proposed by \cite{peterson2026modeling} with applications to \textit{Borrelia burgdorferi} transmission, while the second was a Nicholson-Bailey type host-parasitoid model with intraspecific competition in the host population, to which we added infection dynamics. Although both models exhibit bistability between different types of disease-free attractors, their resulting basic reproductive numbers behave differently. This comparison demonstrates that demographic bistability alone does not necessarily lead to attractor-dependent invasion thresholds; the transmission structure of the system also plays an important role. 

For the tick-host model, evaluating $\mathcal{R}_0$ at different disease-free attractors produced genuinely different invasion thresholds. The $\mathcal{R}_0$ calculated at the positive coexistence fixed point differed from that calculated on the synchronous 2-cycle because the infection experiences different demographic environments along these attractors. At the fixed point, transmission is weighted by the tick host-finding factors $S_{L_\infty}$ and $S_{N_\infty}$. Because these quantities are less than one, they reduce the expected number of successful host-tick contacts and, consequently, the overall transmission potential. In contrast, these terms equal one in the corresponding expression for the synchronous 2-cycle, while host survival appears as the weighting factor because hosts must survive across two time steps to complete the infection pathway.

Even within the $\mathcal{R}_0$ formulation for the synchronous 2-cycle, the closed-form expression depended on the point at which the cycle was initiated. When the system begins with larvae, the infection cycle immediately proceeds from infected hosts to larval ticks, followed by larval development into nymphs and subsequently transmission to new hosts. In contrast, when the system begins with no larvae, this sequence is effectively shifted by one time step, and hosts must survive to the following time step before they can contribute to the infection of larval ticks. Thus, although the two formulations describe the same demographic cycle and, importantly, the same bifurcation threshold, the temporal ordering of infection events influences the resulting expression for the basic reproductive number.

For the parameter set considered, the $\mathcal{R}_0$ calculated on the demographic cycle exceeded that calculated at the positive coexistence fixed point. Notably, varying the demographic reproductive number $\mathcal{R}_d$ produced a region in which $\mathcal{R}_{0_1}<1$ at the fixed point while $\mathcal{R}_{0_2}>1$ on the synchronous 2-cycle. Thus, the same parameters can produce different invasion thresholds depending on the disease-free attractor reached by the population. In this model, the demographic state therefore cannot be disentangled from the interpretation of $\mathcal{R}_0$.

The host-parasitoid model provides a contrasting result. We extended the demographic model by incorporating horizontal and vertical infection transmission and then calculated the basic reproductive number at the positive coexistence fixed point and on the host-only periodic cycles. As with the tick-host model, the periodic reproduction numbers account for the repeated application of the infection and demographic processes over the duration of the cycle. However, after converting the periodic reproduction numbers to a one-step equivalent form, the host component of $\mathcal{R}_0$ was shown to produce the same values at the fixed point and on the host-only cycles. Thus, demographic bistability between a positive coexistence fixed point and a host-only cycle did not, by itself, produce different invasion threshold in this model. 

This result can be understood from the structure of infection transmission in the host-parasitoid model. Infection can be transmitted from infected hosts to parasitoids, but infected parasitoids do not transmit infection back to hosts because parasitized hosts do not survive parasitism. Thus, the infection pathway does not form a complete host--parasitoid--host cycle. At the coexistence fixed point, parasitoids are present, and their vertical transmission contributes an additional component to the fixed-point expression for $\mathcal{R}_0$. On the host-only period cycles, however, parasitoids are absent, so this component is not present. Therefore, any difference in the numerical values of $\mathcal{R}_0$ between the fixed point and the cycles reflects the presence of the parasitoids at the fixed point rather than a difference in the host invasion threshold. Because the parasitoid-only transmission pathway cannot sustain infection independently (since the vertical transmission parameter is less than one), this additional component does not change the invasion threshold.

This distinction illustrates why demographic bistability produces attractor-dependent invasion thresholds in the tick-host model but not in the host-parasitoid model. Specifically, the host-vector (two-population) transmission cycle in the tick-host model leads to difference reproductive numbers on different attractors. In the absence of this transmission cycle, when transmission occurs only directly within a single population, neither model exhibits attractor-dependent reproductive numbers. 

This is illustrated by the two invasion threshold conditions for the host-parasitoid model: 
\begin{align*}
    \ln(\lambda)&=\mu E_\infty+aL_\infty \quad \text{on the fixed point and}\\
    \ln(\lambda)&=\frac{\mu}{k}\sum_{j=1}^k E_{j_\infty} \quad \text{on the $k$-cycle.}
\end{align*}
This forces
\begin{align*}\mu E_\infty+aL_\infty=\frac{\mu}{k}\sum_{j=1}^k E_{j_\infty},
\end{align*}
which are the only differences in the two $\mathcal{R}_0$ expressions and thus show that the two thresholds are necessarily the same. Incorporating transmission from parasitoid to host, however, would change the basic reproduction threshold, as this would make the next generation matrix nontriangular and thus cause two distinct reproductive number thresholds. This comparison highlights the importance of considering host-vector-type transmission cycles, rather than simpler direct transmission, when investigating how demographic bistability influences invasion thresholds. 

Together, these results show that a single invasion threshold does not always adequately characterize invasion when multiple disease-free attractors coexist. However, demographic bistability alone is not sufficient to produce different invasion thresholds. Rather, differences in $\mathcal{R}_0$ arise from the interaction between the demographic attractor and the structure of pathogen transmission. In particular, our results indicate that a true vector-borne transmission pathway, in which infection must pass from host to vector and back to host, is necessary for the demographic attractor to produce different invasion thresholds. Thus, whether a single threshold quantity adequately characterizes invasion depends on the number and type of demographic attractors as well as on how infection moves through the populations within those attractors.  

Some limitations of the current work should be noted. First, our analysis focused on two discrete-time models and therefore does not establish whether the same pattern arises universally in continuous-time systems. Additional types of models should be studied to determine how broadly these results apply. Second, we focused primarily on bistability between a stable fixed point and stable periodic cycles. Other forms of multistability, such as between multiple cycles or chaotic attractors, may lead to additional complexities in the interpretation of $\mathcal{R}_0$. In particular, the periodic next-generation operators considered here hold only on periodic orbits, so nonperiodic attractors may require different techniques.

There are several interesting extensions to this work. Future work could examine whether there are general conditions under which $\mathcal{R}_0$ calculated on a periodic orbit is greater than the value calculated at a fixed point. Another direction is to consider an infection with disease-induced death and investigate how infection feeds back into the demographics after invasion. Specifically, if infection changes population abundance or composition in the long term, the population may move between different demographic attractors as infection becomes established. Understanding these transitions could provide insight into how demographic bistability influences disease invasion as well as long-term disease persistence and prevalence.

Overall, our results demonstrate that the interpretation of the basic reproductive number is closely tied to both the demographic dynamics and the transmission structure of the population. When multiple types of disease-free attractors coexist, attractor-specific reproduction numbers may be necessary to describe invasion. These results emphasize that understanding pathogen invasion in multistable populations requires consideration of both the long-term demographic state and the complete pathway through which infection is transmitted.

\section*{Declarations}

\bmhead{Funding}
This work was supported by the US National Science Foundation under Grant No. DMS-2230790.

\bmhead{Competing Interests}
The authors declare that they have no known competing financial interests or personal relationships that could have appeared to influence the work reported in this paper.

\bmhead{Ethics approval and consent to participate}
Not applicable.

\bmhead{Consent for publication}
Not applicable.

\bmhead{Data availability}
Not applicable.

\bmhead{Materials availability}
Not applicable.

\bmhead{Code availability}
Code used to generate numerical simulations and figures in the text can be found at \url{https://github.com/carlipeterson/bmb-bistability}.

\bmhead{Author Contributions}
Conceptualization, C.P., E.G., C.M.K; 
methodology, C.P., C.M.K.; 
software, C.P.; 
formal analysis, C.P., C.M.K.; 
investigation, C.P., C.M.K.; 
writing---original draft preparation, C.P.; 
writing---review and editing, C.P., C.M.K.; 
visualization, C.P.; 
supervision, E.G., C.M.K.; funding acquisition, C.M.K.

\begin{appendices}

\section{Tick-Host Model}
\subsection{Timeline of Events}\label{ap:timeline}
\begin{itemize}
    \item[1.] Nymphs die (from climate-related factors ($s_N$) and from failure to find a host):
    \begin{align*}
        N_S\left(t+\frac{1}{7}\right)&=s_NN_S\left(t\right)\left(1-e^{-\frac{z_N}{N(t)}\sum_{i=1}^6 b_{i,N}\sigma_{i,N}H_i(t)}\right)\\
        N_I\left(t+\frac{1}{7}\right)&=s_NN_I\left(t\right)\left(1-e^{-\frac{z_N}{N(t)}\sum_{i=1}^6 b_{i,N}\sigma_{i,N}H_i(t)}\right)
    \end{align*}
    \item[2.] Nymphs feed and may become infected or may infect hosts:
    \begin{align*}
        H_{i_S}\left(t+\frac{2}{7}\right)&=H_{i_S}\left(t+\frac{1}{7}\right) e^{-\frac{p_{i,N}\left(t+\frac{1}{7}\right)\alpha_iN_I\left(t+\frac{1}{7}\right)}{H_i\left(t+\frac{1}{7}\right)}}\quad \text{for $i=1,2,3,4,5$}\\
        H_{i_I}\left(t+\frac{2}{7}\right)&=H_{i_I}\left(t+\frac{1}{7}\right)+H_{i_S}\left(t+\frac{1}{7}\right) \left(1-e^{-\frac{p_{i,N}\left(t+\frac{1}{7}\right)\alpha_iN_I\left(t+\frac{1}{7}\right)}{H_i\left(t+\frac{1}{7}\right)}}\right)\quad \text{for $i=1,2,3,4,5$}\\~\\
        N_S\left(t+\frac{2}{7}\right)&=N_S\left(t+\frac{1}{7}\right)\left(1-\sum_{i=1}^5 p_{i,N}\left(t+\frac{1}{7}\right) \beta_{i,N} \frac{H_{i_I}\left(t+\frac{1}{7}\right)}{H_i\left(t+\frac{1}{7}\right)}\right)\\
        N_I\left(t+\frac{2}{7}\right)&=N_I\left(t+\frac{1}{7}\right)+N_S\left(t+\frac{1}{7}\right)\sum_{i=1}^5 p_{i,N}\left(t+\frac{1}{7}\right) \beta_{i,N} \frac{H_{i_I}\left(t+\frac{1}{7}\right)}{H_i\left(t+\frac{1}{7}\right)}
    \end{align*}
    \item[3.] Half of the year's hosts are born for host groups 1--4, larvae hatch, and nymphs molt to adults:
    \begin{align*}
        H_{i_S}\left(t+\frac{3}{7}\right)&=H_{i_S}\left(t+\frac{2}{7}\right)+\frac{\Lambda_i}{2} \quad \text{for $i=1,2,3,4$}\\
         H_{5_S}\left(t+\frac{3}{7}\right)&=H_5\left(t+\frac{2}{7}\right)+\Lambda_5\\
        H_6\left(t+\frac{3}{7}\right)&=H_6\left(t+\frac{2}{7}\right)+\Lambda_6\\~\\
           L_S\left(t+\frac{3}{7}\right) &= \frac{r\frac{A\left(t-\frac{1}{7}\right)}{2}}{\frac{\frac{A\left(t-\frac{1}{7}\right)}{2}}{H_6\left(t-\frac{1}{7}\right)}+K}\\~\\
           A_S\left(t+\frac{3}{7}\right)&=N_S\left(t+\frac{2}{7}\right)\\
            A_I\left(t+\frac{3}{7}\right)&=N_I\left(t+\frac{2}{7}\right)\\~\\
            N_S\left(t+\frac{3}{7}\right)&=0\\
            N_I\left(t+\frac{3}{7}\right)&=0\\
    \end{align*}
    \item[4.] Larvae die (from climate-related factors ($s_L$) and from failure to find a host and feed), and half of the year's host deaths occur:
    \begin{align*}
         L_S\left(t+\frac{4}{7}\right)&=s_LL_S\left(t+\frac{3}{7}\right)\left(1-e^{-\frac{z_L}{L(t)}\sum_{i=1}^6 b_{i,L}H_i\left(t+\frac{3}{7}\right)}\right)\\~\\
        H_{i_S}\left(t+\frac{4}{7}\right) &= H_{i_S}\left(t+\frac{3}{7}\right)e^{-\frac{\mu_{i}}{2}} \quad \text{for $i=1,2,3,4,5$}\\
        H_{i_I}\left(t+\frac{4}{7}\right) &= H_{i_I}\left(t+\frac{3}{7}\right)e^{-\frac{\mu_{i}}{2}} \quad \text{for $i=1,2,3,4,5$}\\
        H_6\left(t+\frac{4}{7}\right)&= H_6\left(t+\frac{3}{7}\right)e^{-\frac{\mu_{6}}{2}}
    \end{align*}
    \item[5.] Larvae feed and potentially become infected, and adults die (from climate-related factors ($s_A$) and from failure to find a host and feed):
    \begin{align*}
        L_S\left(t+\frac{5}{7}\right)&=L_S\left(t+\frac{4}{7}\right)\left(1-\sum_{i=1}^5 p_{i,L}\left(t+\frac{4}{7}\right) \beta_{i,L} \frac{H_{i_I}\left(t+\frac{4}{7}\right)}{H_i\left(t+\frac{4}{7}\right)}\right)\\
        L_I\left(t+\frac{5}{7}\right)&=L_I\left(t+\frac{4}{7}\right)+L_S\left(t+\frac{4}{7}\right)\sum_{i=1}^5 p_{i,L}\left(t+\frac{4}{7}\right) \beta_{i,L} \frac{H_{i_I}\left(t+\frac{4}{7}\right)}{H_i\left(t+\frac{4}{7}\right)}\\~\\
        A_S\left(t+\frac{5}{7}\right)&=s_AA_S\left(t+\frac{4}{7}\right)\left(1-e^{-\frac{z_A}{A(t)} b_{6,A}\sigma_{6,A}H_6\left(t+\frac{4}{7}\right)}\right)\\
        A_I\left(t+\frac{5}{7}\right)&=s_AA_I\left(t+\frac{4}{7}\right)\left(1-e^{-\frac{z_A}{A(t)}b_{6,A}\sigma_{6,A}H_6\left(t+\frac{4}{7}\right)}\right)
    \end{align*}
    \item[6.] Larvae molt to nymphs, female adults feed and lay eggs, and the second half of the year's hosts are born for host groups 1--4:
    \begin{align*}
        N_S\left(t+\frac{6}{7}\right)&=L_S\left(t+\frac{5}{7}\right)\\
        N_I\left(t+\frac{6}{7}\right)&=L_I\left(t+\frac{5}{7}\right)\\~\\
        L_S\left(t+\frac{6}{7}\right)&=0\\
        L_I\left(t+\frac{6}{7}\right)&=0\\~\\
        H_{i_S}\left(t+\frac{6}{7}\right)&=H_{i_S}\left(t+\frac{5}{7}\right)+\frac{\Lambda_i}{2} \quad \text{for $i=1,2,3,4$}\\~\\
        A_S\left(t+\frac{6}{7}\right)&=A_S\left(t+\frac{5}{7}\right)\\
        A_I\left(t+\frac{6}{7}\right)&=A_I\left(t+\frac{5}{7}\right)
    \end{align*}
    \item[7.] All adults die (after reproducing), and the second half of the year's host deaths occur:
    \begin{align*}
        A_S\left(t+1\right)&=0\\
        A_I\left(t+1\right)&=0\\~\\
        H_{i_S}\left(t+1\right) &= H_{i_S}\left(t+\frac{6}{7}\right)e^{-\frac{\mu_{i}}{2}} \quad \text{for $i=1,2,3,4,5$}\\
        H_{i_I}\left(t+1\right) &= H_{i_I}\left(t+\frac{6}{7}\right)e^{-\frac{\mu_{i}}{2}}\quad \text{for $i=1,2,3,4,5$}\\
        H_6\left(t+1\right)&= H_6\left(t+\frac{6}{7}\right)e^{-\frac{\mu_{6}}{2}}
    \end{align*}
\end{itemize}

\newpage

\begin{landscape}

\begin{table}[h!]
\centering
\fontsize{4.75pt}{6pt}\selectfont
\begin{tabular}{l|l|ccccccc}
\toprule
 & &
$t+\frac{1}{7}$ &
$t+\frac{2}{7}$ &
$t+\frac{3}{7}$ &
$t+\frac{4}{7}$ &
$t+\frac{5}{7}$ &
$t+\frac{6}{7}$ &
$t+1$ \\
\midrule

$L_S$ & Event
& 
&
& Larvae hatch
& Larvae die
& \cellcolor{gray!25}\begin{tabular}{c}
Larvae feed\\and get infected\\
(Census)
\end{tabular}
& $L_S\to N_S$
& \\

  & Equation
& $0$
& $0$
& $\frac{r\frac{A\left(t-\frac{4}{7}\right)}{2}}{\frac{\frac{A\left(t-\frac{4}{7}\right)}{2}}{H_6\left(t-\frac{4}{7}\right)}+K}$
& $s_L S_L(t)L_S(t+\frac{3}{7})$
& \cellcolor{gray!25}\begin{tabular}{c}
$L_S(t+\frac{4}{7}) \cdot$\\
$\left(1-\sum_{i=1}^5p_{i,L}\left(t+\frac{4}{7}\right)\beta_{i,L}\frac{H_{i_I}(t+\frac{4}{7})}{H_i(t+\frac{4}{7})}\right)$
\end{tabular}
& $0$
& $0$ \\

\midrule

$L_I$ & Event
& 
&
& 
& Larvae die
& \cellcolor{gray!25}\begin{tabular}{c}
Larvae feed\\and get infected\\
(Census)
\end{tabular}
& $L_I\to N_I$
& \\

  & Equation
& $0$
& $0$
& $0$
& $0$
& \cellcolor{gray!25}\begin{tabular}{c}
$L_I(t+\frac{4}{7})+$\\
$L_S(t+\frac{4}{7}) \cdot$\\
$\left(\sum_{i=1}^5p_{i,L}\left(t+\frac{4}{7}\right)\beta_{i,L}\frac{H_{i_I}(t+\frac{4}{7})}{H_i(t+\frac{4}{7})}\right)$
\end{tabular}
& $0$
& $0$ \\

\midrule

$N_S$ & Event
& Nymphs die
& \cellcolor{gray!25}\begin{tabular}{c}
Nymphs feed\\and get infected\\
(Census)
\end{tabular}
&
$N_S\to A_S$

&
&
& 
$L_S\to N_S$

& \\

  & Equation
& $s_NS_N(t)N_S(t)$
& \cellcolor{gray!25}\begin{tabular}{c}$N_S(t+\frac{1}{7}) \cdot$\\
$\left(1-\sum_{i=1}^5 p_{i,N}\left(t+\frac{1}{7}\right)\beta_{i,N}\frac{H_{i_I}(t+\frac{1}{7})}{H_i(t+\frac{1}{7})}\right)$
\end{tabular}
& $0$
& $0$
& $0$
& $L_S(t+\frac{5}{7})$
& $N_S(t+\frac{6}{7})$\\

\midrule

$N_I$ & Event
& Nymphs die
& \cellcolor{gray!25}\begin{tabular}{c}
Nymphs feed\\and get infected\\
(Census)
\end{tabular}
&
$N_I \to A_I$

&
&
& 
$L_I \to N_I$

& \\

  & Equation
& $s_NS_N(t)N_I(t)$
& \cellcolor{gray!25}\begin{tabular}{c}
$N_I(t+\frac{1}{7})+$\\
$N_S(t+\frac{1}{7})\cdot$\\
$\left(\sum_{i=1}^5 p_{i,N}\left(t+\frac{1}{7}\right)\beta_{i,N}\frac{H_{i_I}(t+\frac{1}{7})}{H_i(t+\frac{1}{7})}\right)$
\end{tabular}
& $0$
& $0$
& $0$
& $L_I(t+\frac{5}{7})$
& $N_I(t+\frac{6}{7})$\\

\midrule

$A_S$ & Event
&
&
& $N_S\to A_S$
&
& \cellcolor{gray!25}\begin{tabular}{c}
Adults die\\
(Census)
\end{tabular}
& \begin{tabular}{c}Female adults\\feed and lay eggs\end{tabular}

& \\

  & Equation
& $0$
& $0$
& $N_S(t+\frac{2}{7})$
& $A_S(t+\frac{3}{7})$
& \cellcolor{gray!25}$s_A S_A(t+\frac{4}{7})A_S(t+\frac{4}{7})$
& $A_S(t+\frac{5}{7})$
& $0$
\\

\midrule

$A_I$ & Event
&
&
& $N_I\to A_I$
&
& \cellcolor{gray!25}\begin{tabular}{c}
Adults die\\
(Census)
\end{tabular}
& \begin{tabular}{c}Female adults feed\\and lay eggs\end{tabular}
& \\

  & Equation
& $0$
& $0$
& $N_I(t+\frac{2}{7})$
& $A_I(t+\frac{3}{7})$
& \cellcolor{gray!25}$s_A S_A(t)A_I(t+\frac{4}{7})$
& $A_I(t+\frac{5}{7})$
& $0$
\\

\bottomrule
\end{tabular}
\caption{Visual representation of how the tick–host dynamics are divided into seven sequential events. Gray cells denote census times where the state variables are defined.}\label{timeline}
\end{table}

\begin{table*}[h!]
\centering
\fontsize{7.5pt}{9pt}\selectfont
\begin{tabular}{l|l|ccccccc}
\toprule
 & &
$t+\frac{1}{7}$ &
$t+\frac{2}{7}$ &
$t+\frac{3}{7}$ &
$t+\frac{4}{7}$ &
$t+\frac{5}{7}$ &
$t+\frac{6}{7}$ &
$t+1$ \\
\midrule
$H_{i_S}$ & Event
& 
& \cellcolor{gray!25}\begin{tabular}{c}
Nymphs feed\\and infect hosts\\
(Census)\\
\end{tabular}
& Hosts born
& Hosts die
& 
& Hosts born
& Hosts die\\

  & Equation
& $H_{i_S}(t)$
& \cellcolor{gray!25}$H_{i_S}(t+\frac{1}{7})e^{-\frac{p_{i,N}\left(t+\frac{1}{7}\right)\alpha_iN_I\left(t+\frac{1}{7}\right)}{H_{i}\left(t+\frac{1}{7}\right)}}$
& $H_{i_S}(t+\frac{2}{7})+\frac{\Lambda_i}{2}$
& $H_{i_S}(t+\frac{3}{7})e^{-\frac{\mu_i}{2}}$
& $H_{i_S}(t+\frac{4}{7})$
& $H_{i_S}(t+\frac{5}{7})+\frac{\Lambda_i}{2}$
& $H_{i_S}(t+\frac{6}{7})e^{-\frac{\mu_i}{2}}$
\\

\midrule

$H_{i_I}$ & Event
& 
& \cellcolor{gray!25}\begin{tabular}{c}
Nymphs feed\\and infect hosts\\
(Census)
\end{tabular}
& 
& Hosts die
& 
& 
& Hosts die\\

  & Equation
& $H_{i_I}(t)$
& \cellcolor{gray!25}\begin{tabular}{c}
$H_{i_I}(t+\frac{1}{7})+$\\
$H_{i_S}(t+\frac{1}{7})\left(1-e^{-\frac{p_{i,N}\left(t+\frac{1}{7}\right)\alpha_iN_I\left(t+\frac{1}{7}\right)}{H_{i}\left(t+\frac{1}{7}\right)}}\right)$
\end{tabular}
& $H_{i_I}(t+\frac{2}{7})$
& $H_{i_I}(t+\frac{3}{7})e^{-\frac{\mu_i}{2}}$
& $H_{i_I}(t+\frac{4}{7})$
& $H_{i_I}(t+\frac{5}{7})$
& $H_{i_I}(t+\frac{6}{7})e^{-\frac{\mu_i}{2}}$
\\

\midrule

$H_{5_S}$ & Event
& 
& \cellcolor{gray!25}\begin{tabular}{c}
Nymphs feed\\and infect hosts\\
(Census)
\end{tabular}
& Hosts born
& Hosts die
& 
&
& Hosts die\\

  & Equation
& $H_{5_S}(t)$
& \cellcolor{gray!25}$H_{5_S}(t+\frac{1}{7})e^{-\frac{p_{5,N}\left(t+\frac{1}{7}\right)\alpha_5N_I\left(t+\frac{1}{7}\right)}{H_{5}\left(t+\frac{1}{7}\right)}}$
& $H_{5_S}(t+\frac{2}{7})+\Lambda_5$
& $H_{5_S}(t+\frac{3}{7})e^{-\frac{\mu_5}{2}}$
& $H_{5_S}(t+\frac{4}{7})$
& $H_{5_S}(t+\frac{5}{7})$
& $H_{5_S}(t+\frac{6}{7})e^{-\frac{\mu_5}{2}}$
\\

\midrule

$H_{5_I}$ & Event
& 
& \cellcolor{gray!25}\begin{tabular}{c}
Nymphs feed\\and infect hosts\\
(Census)
\end{tabular}
&
& Hosts die
& 
&
& Hosts die\\

  & Equation
& $H_{5_I}(t)$
& \cellcolor{gray!25}\begin{tabular}{c} $H_{5_I}(t+\frac{1}{7})+$\\
$H_{5_S}(t+\frac{1}{7})\left(1-e^{-\frac{p_{5,N}\left(t+\frac{1}{7}\right)\alpha_5N_I\left(t+\frac{1}{7}\right)}{H_{5}\left(t+\frac{1}{7}\right)}}\right)$
\end{tabular}
& $H_{5_I}(t+\frac{2}{7})$
& $H_{5_I}(t+\frac{3}{7})e^{-\frac{\mu_5}{2}}$
& $H_{5_I}(t+\frac{4}{7})$
& $H_{5_I}(t+\frac{5}{7})$
& $H_{5_I}(t+\frac{6}{7})e^{-\frac{\mu_5}{2}}$
\\

\midrule

$H_6$ & Event
&
&
& Hosts born
& Hosts die
& \cellcolor{gray!25}(Census)
&
& Hosts die\\

  & Equation
& $H_6(t)$
& $H_6(t+\frac{1}{7})$
& $H_6(t+\frac{2}{7})+\Lambda_6$
& $H_6(t+\frac{3}{7})e^{-\frac{\mu_6}{2}}$
& \cellcolor{gray!25}$H_6(t+\frac{4}{7})$
& $H_6(t+\frac{5}{7})$
& $H_6(t+\frac{6}{7})e^{-\frac{\mu_6}{2}}$
\\
\bottomrule
\end{tabular}
\caption*{Table~\ref{timeline} continued.}
\end{table*}

\end{landscape}

\subsection{Basic Reproductive Number Derivation - Constant Population}\label{ap:BRN}
To derive the basic reproductive number when populations approach a stable fixed point (denoted here by $\mathcal{R}_{0_1}$), we linearize the infection subsystem of Model \eqref{fullsystem} around the disease-free fixed point. The infection subsystem is given by
\begin{equation}
\begin{aligned}
    L_I\left(t+1\right)&=\frac{r\frac{A_{\infty}}{2}}{\frac{\frac{A_{\infty}}{2}}{H_{6_{\infty}}}+K}s_LS_{L_{\infty}}\left(\sum_{i=1}^5 p_{{i,L}_{\infty}} \beta_{i,L} \frac{H_{i_I}\left(t\right)}{H_{i_{\infty}}+\frac{\Lambda_i}{k_i}}\right)\\
    N_I\left(t+1\right)&=s_NS_{N_\infty}\left(L_I\left(t\right)+L_S\left(t\right)\left(\sum_{i=1}^5p_{{i,N}_\infty}\beta_{i,N}\frac{H_{i_I}\left(t\right)e^{-\mu_i}}{H_{i_\infty}}\right)\right)\\
    A_I\left(t+1\right)&=s_AS_{A_\infty}s_NS_{N_\infty}\left(L_I\left(t\right)+L_S\left(t\right)\left(\sum_{i=1}^5p_{{i,N}_\infty}\beta_{i,N}\frac{H_{i_I}\left(t\right)e^{-\mu_i}}{H_{i_\infty}}\right)\right)\\
    H_{i_I}\left(t+1\right)&=\left(H_{i_I}\left(t\right)+ \left(H_{i_S}(t)+\lambda_i\right)\left(1-e^{-\frac{p_{{i,N}_\infty}\alpha_{i}s_NS_{N_\infty}L_I\left(t\right)}{H_{i_\infty}}}\right)\right)e^{-\mu_i},
\end{aligned}
\end{equation}
where
\begin{align*}
    \lambda_i = \frac{\Lambda_i}{2}\left(1+e^{\frac{\mu_i}{2}}\right) \quad \text{for $i \in \{1,2,3,4\}$}, \quad
    \lambda_5 = \Lambda_5.
\end{align*}
At the disease-free fixed point, all hosts are susceptible and tick survival and contact parameters take their equilibrium values $S_{j_\infty}$ and $p_{{i,j}_{\infty}}$. 

Because the infected larvae and host equations do not depend on $N_I$ or $A_I$, these equations decouple, and we can analyze the larvae-host subsystem separately. 

The Jacobian matrix of this system evaluated at the disease-free fixed point is
\begin{align*}
    J = \begin{bmatrix}
    \vspace{0.2cm}
        0&R_{L1}&R_{L2}&R_{L3}&R_{L4}&R_{L5}\\
        \vspace{0.2cm}
        R_{1L}e^{-\mu_1}&e^{-\mu_1}&0&0&0&0\\
        \vspace{0.2cm}
        R_{2L}e^{-\mu_2}&0&e^{-\mu_2}&0&0&0\\
        \vspace{0.2cm}
        R_{3L}e^{-\mu_3}&0&0&e^{-\mu_3}&0&0\\
        \vspace{0.2cm}
        R_{4L}e^{-\mu_4}&0&0&0&e^{-\mu_4}&0\\
        \vspace{0.2cm}
        R_{5L}e^{-\mu_5}&0&0&0&0&e^{-\mu_5}
    \end{bmatrix}
\end{align*}
where
\begin{align*}
    R_{Li}&=\frac{r\frac{A_{\infty}}{2}}{\frac{\frac{A_{\infty}}{2}}{H_{6_{\infty}}}+K}s_LS_{L_\infty} \frac{p_{{i,L}_\infty}\beta_{i,L}}{H_{i_{\infty}}+\frac{\Lambda_i}{k_i}},\\
    R_{iL}&=p_{{i,N}_\infty}\alpha_is_NS_{N_\infty}\frac{\left(H_{i_\infty}+\lambda_i\right)}{H_{i_\infty}}.
\end{align*}
The quantity $R_{Li}R_{iL}$ represents one complete larva-host-larva transmission loop involving host group $i$. Summing across host groups therefore accounts for the expected number of infected larvae produced by a larva that survives through the tick life cycle.

Following the next-generation operator method for discrete-time systems \citep{allen2008basic}, we split the Jacobian into $F$ and $T$ matrices. Here, $F$ contains the terms associated with new infections. The resulting matrix is
\begin{align*}F=
    \begin{bmatrix}
    \vspace{0.2cm}
        0&R_{L1}&R_{L2}&R_{L3}&R_{L4}&R_{L5}\\
        \vspace{0.2cm}
        R_{1L}e^{-\mu_1}&0&0&0&0&0\\
        \vspace{0.2cm}
        R_{2L}e^{-\mu_2}&0&0&0&0&0\\
        \vspace{0.2cm}
        R_{3L}e^{-\mu_3}&0&0&0&0&0\\
        \vspace{0.2cm}
        R_{4L}e^{-\mu_4}&0&0&0&0&0\\
        \vspace{0.2cm}
        R_{5L}e^{-\mu_5}&0&0&0&0&0
    \end{bmatrix},
\end{align*}

The transition matrix $T$ captures host turnover and is given by
\begin{align*}
    T=\begin{bmatrix}
        0&0&0&0&0&0\\
        0&e^{-\mu_1}&0&0&0&0\\
        0&0&e^{-\mu_2}&0&0&0\\
        0&0&0&e^{-\mu_3}&0&0\\
        0&0&0&0&e^{-\mu_4}&0\\
        0&0&0&0&0&e^{-\mu_5}
    \end{bmatrix}.
\end{align*}
The next generation matrix is therefore
\begin{align*}
    Q=F(I-T)^{-1},
\end{align*}
where $I$ is the $6\times 6$ identity matrix.

The basic reproductive number $\mathcal{R}_{0_1}$ is the spectral radius of $Q$:
\begin{align*}
    \mathcal{R}_{0_1}=\rho(Q).
\end{align*}
Solving for the dominant eigenvalue yields
\begin{align*}
    \mathcal{R}_{0_1} &=  \sqrt{
        \sum_{i=1}^5 R_{Li}R_{iL}\frac{e^{-\mu_i}}{1-e^{-\mu_i}}}.
\end{align*}

Solving the inequality $\mathcal{R}_{0_1}<1$ yields
\begin{align*}
    \sum_{i=1}^5R_{iL}R_{Li}\frac{e^{-\mu_i}}{1-e^{-\mu_i}}<1.
\end{align*}
The magnitude of $\mathcal{R}_{0_1}$ therefore depends on two interacting mechanisms: the efficiency of the larva-host-larva transmission cycle and the probability that hosts survive long enough to participate in repeated transmission events.

Now let $\displaystyle R_i=\frac{R_{iL}R_{Li}}{S_{L_\infty}S_{N_\infty}}$. Using this definition, we can express the above threshold in the form
\begin{align*}
    \hat{\mathcal{R}}_{0_1}=\sum_{i=1}^5R_iS_{L_\infty}S_{N_\infty}\frac{e^{-\mu_i}}{1-e^{-\mu_i}}<1,
\end{align*}
The quantity $\hat{\mathcal{R}}_{0_1}$ is not itself equivalent to $\mathcal{R}_{0_1}$, but it provides an equivalent invasion threshold:
\begin{align*}
    \mathcal{R}_{0_1} <1 \iff \hat{\mathcal{R}}_{0_1}<1.
\end{align*}
This alternative expression will be useful for comparison with the threshold conditions obtained for oscillatory population dynamics (see Sections~\ref{ap:BRN_oscillate} and~\ref{BRN_2}). 

Consequently, the disease-free fixed point is LAS whenever either inequality is satisfied. In this case, the combined contribution of larva-host-larva transmission loops is insufficient to overcome losses in the tick life cycle.

\subsection{Basic Reproductive Number Derivation - Oscillating Populations}\label{ap:BRN_oscillate}
Following the periodic next-generation approach of \citep{van2019demographic}, we compute the basic reproductive number when population sizes do not converge to a fixed point but instead approach a stable 2-cycle. In this setting, demographic variables alternate between two values corresponding to the two points of the cycle. The basic reproductive number is therefore obtained by linearizing the infection subsystem over one full period, consisting of two one-year time steps.

To derive $\mathcal{R}_0$ for the synchronous 2-cycle, let $J_1$ and $J_2$ denote the Jacobian matrices of the infection subsystem evaluated at the two points of the cycle: 
\begin{align*}
    (L_1,N_1,A_1) &= \left(\frac{2H_{6_\infty}(rS-K)}{s_Ns_A},0,0\right), \\(L_2,N_2,A_2) &= \left(0,\frac{2H_{6_\infty}(rS-K)}{s_A},2H_{6_\infty}(rS-K)\right).
\end{align*}
Because the ordering of the two states in the cycle depends on initial conditions, either point may occur first in the sequence. We therefore consider both possibilities.

Suppose first that $(L_1,N_1,A_1)$ precedes $(L_2,N_2,A_2)$. The Jacobians evaluated at these points are
\begin{align*}
    J_1=\begin{bmatrix}
        0&0&0&0&0&0\\
        \frac{R_{1L}e^{-\mu_1}}{S_{N_\infty}}&e^{-\mu_1}&0&0&0&0\\
        \frac{R_{2L}e^{-\mu_2}}{S_{N_\infty}}&0&e^{-\mu_2}&0&0&0\\
        \frac{R_{3L}e^{-\mu_3}}{S_{N_\infty}}&0&0&e^{-\mu_3}&0&0\\
        \frac{R_{4L}e^{-\mu_4}}{S_{N_\infty}}&0&0&0&e^{-\mu_4}&0\\
        \frac{R_{5L}e^{-\mu_5}}{S_{N_\infty}}&0&0&0&0&e^{-\mu_5}
    \end{bmatrix}
\end{align*}
and 
\begin{align*}
    J_2=\begin{bmatrix}
        0&\frac{R_{L1}}{S_{L_\infty}}&\frac{R_{L2}}{S_{L_\infty}}&\frac{R_{L3}}{S_{L_\infty}}&\frac{R_{L4}}{S_{L_\infty}}&\frac{R_{L5}}{S_{L_\infty}}\\
        R_{1L}e^{-\mu_1}&e^{-\mu_1}&0&0&0&0\\
        R_{2L}e^{-\mu_2}&0&e^{-\mu_2}&0&0&0\\
        R_{3L}e^{-\mu_3}&0&0&e^{-\mu_3}&0&0\\
        R_{4L}e^{-\mu_4}&0&0&0&e^{-\mu_4}&0\\
        R_{5L}e^{-\mu_5}&0&0&0&0&e^{-\mu_5}
    \end{bmatrix},
\end{align*}
where $J_1$ is evaluated at $(L_1,N_1,A_1)$ and $J_2$ is evaluated at $(L_2,N_2,A_2)$. The quantities $R_{iL}$ and $R_{Li}$ are as defined in Appendix~\ref{ap:BRN}.

To capture infection over one complete cycle, we consider the product $J_2J_1$ and decompose it as
\begin{align*}
    J_2J_1 = F+T,
\end{align*}
where $F$ represents new infections and $T$ represents demographic transitions. As in the constant-population (fixed points) case, this yields the block structure
\begin{align*}F=
    \begin{bmatrix}
        \vspace{0.3cm}
         \sum_{i=1}^5\frac{R_{iL}R_{Li}e^{-\mu_i}}{S_{L_\infty}S_{N_\infty}}&\frac{R_{L1}e^{-\mu_1}}{S_{L_\infty}}&\frac{R_{L2}e^{-\mu_2}}{S_{L_\infty}}&\frac{R_{L3}e^{-\mu_3}}{S_{L_\infty}}&\frac{R_{L4}e^{-\mu_4}}{S_{L_\infty}}&\frac{R_{L5}e^{-\mu_5}}{S_{L_\infty}}\\
        \vspace{0.3cm}
        \frac{R_{1L}e^{-2\mu_1}}{S_{N_\infty}}&0&0&0&0&0\\
        \vspace{0.3cm}
        \frac{R_{2L}e^{-2\mu_2}}{S_{N_\infty}}&0&0&0&0&0\\
        \vspace{0.3cm}
        \frac{R_{3L}e^{-2\mu_3}}{S_{N_\infty}}&0&0&0&0&0\\
        \vspace{0.3cm}
        \frac{R_{4L}e^{-2\mu_4}}{S_{N_\infty}}&0&0&0&0&0\\
        \frac{R_{5L}e^{-2\mu_5}}{S_{N_\infty}}&0&0&0&0&0
    \end{bmatrix}.
\end{align*}

Survival over two time steps is described by
\begin{align*}
    T = \begin{bmatrix}
        0&0&0&0&0&0\\
        0&e^{-2\mu_1}&0&0&0&0\\
        0&0&e^{-2\mu_2}&0&0&0\\
        0&0&0&e^{-2\mu_3}&0&0\\
        0&0&0&0&e^{-2\mu_4}&0\\
        0&0&0&0&0&e^{-2\mu_5}
    \end{bmatrix}.
\end{align*}
The next-generation matrix over one full cycle is therefore
\begin{align*}
    Q = F(I-T)^{-1},
\end{align*}
and the basic reproductive number is the spectral radius
\begin{align*}
    \mathcal{R}_{0_2}=\rho(Q).
\end{align*}
Noting that in this case $S_{L_\infty}=S_{N_\infty}=1$ and solving for the dominant eigenvalue gives
\begin{align*}
    \mathcal{R}_{0_2} &= \frac{1}{2} \left(\sum_{i=1}^5R_{iL}R_{Li}e^{-\mu_i} + \sqrt{
        \left(\sum_{i=1}^5R_{iL}R_{Li}e^{-\mu_i}\right)^2 + 4 \sum_{i=1}^5 R_{Li}R_{iL}\frac{e^{-3\mu_i}}{1-e^{-2\mu_i}}}\right).
\end{align*}
As in the constant-population case, $R_{Li}R_{iL}$ corresponds to a complete larva-host-larva infection loop. Different from the fixed-point version, this expression contains an additional factor of $e^{-2\mu_i}$, reflecting the loss of hosts over two time steps. The denominator $1-e^{-2\mu_i}$ arises because infectious hosts must survive across both phases of the 2-cycle before contributing again to larval infection.

Using 
\begin{align*}
    R_i = \frac{R_{iL}R_{Li}}{S_{L_\infty}S_{N_\infty}},\quad i \in \{1,2,3,4,5\},
\end{align*}
we obtain
\begin{align*}
    \mathcal{R}_{0_2} &= \frac{1}{2} \left(\sum_{i=1}^5R_ie^{-\mu_i} + \sqrt{
        \left(\sum_{i=1}^5R_ie^{-\mu_i}\right)^2 + 4 \sum_{i=1}^5R_ie^{-\mu_i}\frac{e^{-2\mu_i}}{1-e^{-2\mu_i}}}\right).
\end{align*}

Simplifying $\mathcal{R}_{0_2}<1$ yields the threshold condition
\begin{align*}
    \sum_{i=1}^5R_i\frac{1}{1+e^{-\mu_i}}\frac{e^{-\mu_i}}{1-e^{-\mu_i}}<1,
\end{align*}
which ensures that infection cannot replace itself over one full demographic cycle, and hence the disease-free 2-cycle is locally asymptotically stable. 

Now suppose the ordering is reversed, so that $J_1$ is evaluated at $(L_2,N_2,A_2)$ and $J_2$ at $(L_1,N_1,A_1)$. Repeating the above steps yields
\begin{align*}
    \hat{\mathcal{R}}_{0_2}= \sum_{i=1}^5 R_{Li}R_{iL}\frac{e^{-\mu_i}}{1-e^{-2\mu_i}}= \sum_{i=1}^5 R_i\frac{e^{-\mu_i}}{1-e^{-2\mu_i}}.
\end{align*}
Although this expression differs algebraically from the previous formulation of $\mathcal{R}_{0_2}$, solving $\hat{\mathcal{R}}_{0_2}<1$ leads to the same threshold condition
\begin{align*}
    \sum_{i=1}^5R_i\frac{1}{1+e^{-\mu_i}}\frac{e^{-\mu_i}}{1-e^{-\mu_i}}<1.
\end{align*}
Thus, while the two formulations yield different numerical expressions for the 2-cycle basic reproductive number, they are equivalent for determining the invasion threshold.

The asynchronous 2-cycle case leads to substantially more complicated Jacobian products, and a closed form expression for the basic reproductive number is not easily obtained. We therefore evaluate $\mathcal{R}_{0_2}$ numerically in that setting (see Section~\ref{quantitative_analysis}).

\subsection{Existence of Endemic Fixed Point}\label{ap:ee}
We now state a necessary and sufficient condition for the existence of a unique endemic fixed point of Model~\eqref{fullsystem}.
\begin{theorem}
Model~\eqref{fullsystem} has a unique endemic fixed point if and only if the positive coexistence fixed point is locally asymptotically stable (i.e., $\mathcal{R}_d>1$) and $\mathcal{R}_{0_1}>1$.
\end{theorem}
\begin{proof}
We first reduce Model~\eqref{fullsystem} to a single equation for the infected larval population. Assuming all other populations are at equilibrium, we express $H_{i_I}$ as functions of $L_I$ and substitute into the $L_I$ equation to obtain
\begin{align*}
    L_I=f(L_{I}),
\end{align*}
where 
\begin{align*}
    f(L_I)=\frac{r\frac{A_{\infty}}{2}}{\frac{\frac{A_{\infty}}{2}}{H_{6_{\infty}}}+K}s_LS_{L_\infty}\left(\sum_{i=1}^5 p_{{i,L}_{\infty}} \beta_{i,L} \frac{H_{i_{\infty}}+\lambda_i}{H_{i_\infty}+\frac{\Lambda_i}{k_i}}\left(\frac{\left(1-e^{-\frac{p_{{i,N}_\infty}\alpha_{i}s_NS_{N_\infty}L_{I}}{H_{i_\infty}}}\right)e^{-\mu_i}}{1-e^{-\frac{p_{{i,N}_\infty}\alpha_{i}s_NS_{N_\infty}L_{I}}{H_{i_\infty}}}e^{-\mu_i}}\right)\right).
\end{align*}
The function $f: \mathbb{R}_{\ge0}\to\mathbb{R}$ is continuous and satisfies $f(0)=0$. 

Differentiation shows that $f'(L_I)>0$ and $f''(L_I)<0$ for all $L_I \geq 0$, implying that $f$ is strictly increasing and concave down.

Evaluating the derivative at the origin yields 
\begin{align*}
f'(0)=\frac{r\frac{A_{\infty}}{2}}{\frac{\frac{A_{\infty}}{2}}{H_{6_{\infty}}}+K}s_LS_{L_\infty}s_NS_{N_\infty}\left(\sum_{i=1}^5 p_{{i,L}_{\infty}}p_{{i,N}_{\infty}}\alpha_i \beta_{i,L}\frac{H_{i_\infty}+\lambda_i}{H_{i_\infty}+\frac{\Lambda_i}{k_i}}\frac{e^{-\mu_i}}{1-e^{-\mu_i}}\right)=\hat{\mathcal{R}}_{0_1}.
\end{align*}
Moreover,
\begin{align*}
    \lim_{L_I \to \infty} f(L_I)<\infty.
\end{align*}
Consequently, since $f$ is increasing, concave down, and satisfies $f(0)=0$, a unique positive endemic fixed point exists if and only if $f'(0)>1$, i.e.,
\begin{align*}
    \hat{\mathcal{R}}_{0_1}>1.
\end{align*}
Because $f'(L_I)>0$, the condition $f'(L_{I_{\infty}}),1$ implies $|f'(L_{I_\infty})|<1$, establishing local asymptotic stability.
\end{proof}

\section{Host-Parasitoid Model}\label{qualanalysissystem2}

\subsection{Timeline of Events}\label{ap:timeline2}
\begin{itemize}
    \item[1.] Hosts and parasitoids are born
    \begin{align*}
        E_S\left(t+\frac{1}{5}\right) &= 0\\
        E_I\left(t+\frac{1}{5}\right) &= 0\\
        H_S\left(t+\frac{1}{5}\right) & = E_S\left(t-\frac{1}{5}\right)\\
        H_I\left(t+\frac{1}{5}\right) & = E_I\left(t-\frac{1}{5}\right)\\
        L_S\left(t+\frac{1}{5}\right) &= 0\\
        L_I\left(t+\frac{1}{5}\right) &= 0\\
        P_S\left(t+\frac{1}{5}\right) &= L_S\left(t-\frac{1}{5}\right)\\
        P_I\left(t+\frac{1}{5}\right) &= L_I\left(t-\frac{1}{5}\right)\\
    \end{align*}
    \item[2.] Hosts infect other hosts
    \begin{align*}
        H_S\left(t+\frac{2}{5}\right) & = H_S\left(t+\frac{1}{5}\right)\left(1-\beta \frac{H_I\left(t+\frac{1}{5}\right)}{H\left(t+\frac{1}{5}\right)}\right)\\
        H_I\left(t+\frac{1}{5}\right) & = H_I\left(t+\frac{1}{5}\right)+H_S\left(t+\frac{1}{5}\right)\left(\beta \frac{H_I\left(t+\frac{1}{5}\right)}{H\left(t+\frac{1}{5}\right)}\right)
    \end{align*}
    \item[3.] Hosts lay eggs
    \begin{align*}
        E_S\left(t+\frac{3}{5}\right) &= \lambda \left(H_S\left(t+\frac{2}{5}\right)+(1-v)H_I\left(t+\frac{2}{5}\right)\right)e^{-\mu H\left(t+\frac{2}{5}\right)}\\
        E_I\left(t+\frac{3}{5}\right) &= \lambda vH_I\left(t+\frac{2}{5}\right)e^{-\mu H\left(t+\frac{2}{5}\right)}
    \end{align*}
    \item[4.] Larvae infest host eggs and larvae get infected from hosts
    \begin{align*}
        E_S\left(t+\frac{4}{5}\right) &= E_S\left(t+\frac{3}{5}\right)e^{-aP\left(t+\frac{3}{5}\right)}\\
        E_I\left(t+\frac{4}{5}\right) &= E_I\left(t+\frac{3}{5}\right)e^{-aP\left(t+\frac{3}{5}\right)}\\
        L_S\left(t+\frac{4}{5}\right)&=\frac{(1-w)P_I\left(t+\frac{3}{5}\right)+P_S\left(t+\frac{3}{5}\right)}{P\left(t+\frac{3}{5}\right)}b\left(E_S\left(t+\frac{3}{5}\right)+(1-q)E_I\left(t+\frac{3}{5}\right)\right)\left(1-e^{-aP\left(t+\frac{3}{5}\right)}\right)\\
        L_I\left(t+\frac{4}{5}\right)&=\bigg(\frac{wP_I\left(t+\frac{3}{5}\right)}{P\left(t+\frac{3}{5}\right)}bE\left(t+\frac{3}{5}\right)+\frac{P_S\left(t+\frac{3}{5}\right)+(1-w)P_I\left(t+\frac{3}{5}\right)}{P\left(t+\frac{3}{5}\right)}b\left(qE_I\left(t+\frac{3}{5}\right)\right)\bigg)\cdot\\
        &\left(1-e^{-aP\left(t+\frac{3}{5}\right)}\right)
    \end{align*}
    \item[5.] Hosts and parasitoids die
        \begin{align*}
            H_S\left(t+1\right) & = 0\\
        H_I\left(t+1\right) & = 0\\
        P_S\left(t+1\right) &= 0\\
        P_I\left(t+1\right) &= 0
        \end{align*}
\end{itemize}

\begin{landscape}
{
\renewcommand{\arraystretch}{1.6}
\begin{table}[!ht]
\centering
\fontsize{5pt}{5.6pt}\selectfont
\begin{tabular}{l|l|ccccc}
\toprule
 & &
$t+\frac{1}{5}$ &
$t+\frac{2}{5}$ &
$t+\frac{3}{5}$ &
$t+\frac{4}{5}$ &
$t+1$ \\
\midrule
$E_S$ & Event & & & Hosts lay eggs & \cellcolor{gray!25}\begin{tabular}{c}Larvae infest eggs\\(Census)\end{tabular} & \\
& Eqn & $0$ & $0$ & \begin{tabular}{c}$\lambda \left(H_S\left(t+\frac{2}{5}\right)+(1-v)H_I\left(t+\frac{2}{5}\right)\right)\cdot$\\$e^{-\mu H\left(t+\frac{2}{5}\right)}$\end{tabular} & \cellcolor{gray!25}$E_S\left(t+\frac{3}{5}\right)e^{-aP\left(t+\frac{3}{5}\right)}$ & $E_S\left(t+\frac{4}{5}\right)$\\
\midrule
$E_I$ & Event & & & Hosts lay eggs & \cellcolor{gray!25}Larvae infest eggs & \\
& Eqn & $0$ & $0$ & $\lambda vH_I\left(t+\frac{2}{5}\right) e^{-\mu H\left(t+\frac{2}{5}\right)}$ & \cellcolor{gray!25}$E_I\left(t+\frac{3}{5}\right)e^{-aP\left(t+\frac{3}{5}\right)}$ & $E_I\left(t+\frac{4}{5}\right)$\\
\midrule
$H_S$ & Event
& Hosts are born
& Hosts infect each other
& 
& \cellcolor{gray!25}
& Hosts die
\\

& Eqn
& $E_S\left(t-\frac{1}{5}\right)$
& $H_S\left(t+\frac{1}{5}\right)\left(1-\beta \frac{H_I\left(t+\frac{1}{5}\right)}{H\left(t+\frac{1}{5}\right)}\right)$
& $H_S\left(t+\frac{2}{5}\right)$
& \cellcolor{gray!25}$H_S\left(t+\frac{3}{5}\right)$
& $0$\\

\midrule

$H_I$ & Event
& Hosts are born
& Hosts infect each other
&
&\cellcolor{gray!25}
& Hosts die
\\

  & Eqn
& $E_I\left(t-\frac{1}{5}\right)$
& \begin{tabular}{c}$H_I\left(t+\frac{1}{5}\right)+$\\$H_S\left(t+\frac{1}{5}\right)\left(\beta \frac{H_I\left(t+\frac{1}{5}\right)}{H\left(t+\frac{1}{5}\right)}\right)$\end{tabular}
& $H_I\left(t+\frac{2}{5}\right)$
& \cellcolor{gray!25}$H_I\left(t+\frac{3}{5}\right)$
& $0$
\\

\midrule

$L_S$ & Event
& & & & \cellcolor{gray!25}\begin{tabular}{c}Larvae infest host eggs and get infected\\(Census)\end{tabular} &\\
& Eqn
& $0$
& $0$
& $0$
& \cellcolor{gray!25}\begin{tabular}{c}$\frac{(1-w)P_I\left(t+\frac{3}{5}\right)+P_S\left(t+\frac{3}{5}\right)}{P\left(t+\frac{3}{5}\right)}\cdot$\\
$b\left(E_S\left(t+\frac{3}{5}\right)+(1-q)E_I\left(t+\frac{3}{5}\right)\right)\left(1-e^{-aP\left(t+\frac{3}{5}\right)}\right)$\end{tabular}
& $L_S\left(t+\frac{3}{5}\right)$\\
\midrule
$L_I$ & Event
& & & & \cellcolor{gray!25}\begin{tabular}{c}Larvae infest host eggs and get infected\\(Census)\end{tabular} &\\
& Eqn
& $0$
& $0$
& $0$
& \cellcolor{gray!25}\begin{tabular}{c}$\bigg(\frac{wP_I\left(t+\frac{3}{5}\right)}{P\left(t+\frac{3}{5}\right)}bE\left(t+\frac{3}{5}\right)+$\\
$\frac{P_S\left(t+\frac{3}{5}\right)+(1-w)P_I\left(t+\frac{3}{5}\right)}{P\left(t+\frac{3}{5}\right)}b\left(qE_I\left(t+\frac{3}{5}\right)\right)\bigg)\cdot$\\
$\left(1-e^{-aP\left(t+\frac{3}{5}\right)}\right)$\end{tabular}
& $L_S\left(t+\frac{3}{5}\right)$\\
\midrule
$P_S$ & Event
& Parasites are born
& 
& 
&\cellcolor{gray!25}
& Parasitoids die
\\

  & Eqn
& $L_S\left(t-\frac{1}{5}\right)$
& $P_S\left(t+\frac{1}{5}\right)$
& $P_S\left(t+\frac{2}{5}\right)$
& \cellcolor{gray!25}$P_S\left(t+\frac{3}{5}\right)$
& $0$
\\
\midrule
$P_I$ & Event
& Parasites are born
& 
& 
&\cellcolor{gray!25}
& Parasitoids die
\\

  & Eqn
& $L_I\left(t-\frac{1}{5}\right)$
& $P_I\left(t+\frac{1}{5}\right)$
& $P_I\left(t+\frac{2}{5}\right)$
& \cellcolor{gray!25}$P_I\left(t+\frac{3}{5}\right)$
& $0$
\\
\bottomrule
\end{tabular}

\caption{Visual representation of how the host-parasitoid dynamics are divided into sequential events. Gray cells denote census times where the state variables are defined.}
\label{timeline2}
\end{table}
} 
\end{landscape}
\subsection{Fixed-Point Stability}\label{ap:fp_stability2}

The demographic system given in Section~\ref{modeldevelop2} is given by
\begin{equation}
    \begin{aligned}
        E(t+1) &= \lambda E(t)e^{-\mu E(t)}e^{-aL(t)}\\
        L(t+1) &= b\lambda E(t)e^{-\mu E(t)}(1-e^{-aL(t)}).
    \end{aligned}
\end{equation}
The fixed points of this system are
\begin{align*}
    E_{00} &= (0,0)\\
    E_{+0} &= \left(\frac{\ln(\lambda)}{\mu},0\right)\\
    E_{++} &= (E_{\infty},L_{\infty}),
\end{align*}
where the positive coexistence fixed point satisfies 
\begin{align*}
    E_{\infty} = \begin{cases}
        \frac{L_{\infty}}{b\lambda e^{-\mu E_{\infty}}(1-e^{-aL_{\infty}})},\\
        \frac{\ln(\lambda)-aL_{\infty}}{\mu}
    \end{cases}.
\end{align*}
The extinction fixed point ($E_{00}$) always exists, while the host-only fixed point ($E_{+0}$) exists for $\lambda>1$. The positive coexistence fixed point exists for $\lambda>e^{\frac{\mu}{ab}}$ (see Appendix~\ref{ap:cofp2} for details).

The Jacobian matrix of Model~\eqref{system2noinfect} is
\begin{align*}
    \begin{bmatrix}
        \lambda e^{-\mu E_{\infty}}e^{-aL_{\infty}}(1-\mu E_{\infty}) & -a\lambda e^{-\mu E_{\infty}}e^{-aL_{\infty}}E_{\infty}\\
        b\lambda e^{-\mu E_{\infty}}(1-e^{-aL_{\infty}})(1-\mu E_{\infty}) & abe^{-\mu E_{\infty}}e^{-aL_{\infty}}E_{\infty}
    \end{bmatrix}.
\end{align*}

Evaluating the Jacobian at ($E_{00}$) gives
    \begin{align*}
        J(E_{00}) = 
        \begin{bmatrix}
            \lambda & 0 \\
            0 & 0 
        \end{bmatrix}
    \end{align*}

The eigenvalues are $\lambda_1=0$ and $\lambda_2=\lambda$, and therefore $E_{00}$ is LAS for $\lambda<1$. 

Note that this analysis also shows that $\mathcal{R}_d=\lambda$, since this value determines persistence or extinction of the host (and consequently the parasitoid) population. 

Evaluating the Jacobian at the host-only fixed point ($E_{+0}$) gives
\begin{align*}
    J(E_{+0}) = \begin{bmatrix}
        1-\ln(\lambda) & -\frac{a \ln(\lambda)}{\mu}\\
        0 & \frac{ab\ln(\lambda)}{\mu}.
    \end{bmatrix}
\end{align*}
Here, the eigenvalues are $\lambda_1 = 1-\ln(\lambda)$ and $\lambda_2 = \frac{ab\ln(\lambda)}{\mu}$. Therefore, $E_{+0}$ is LAS if and only if
\begin{align*}
    |1-\ln(\lambda)|<1 \quad \text{and} \quad \frac{ab\ln(\lambda)}{\mu}<1,
\end{align*}
or equivalently,
\begin{align*}
    1<\lambda<\min\left\{e^2, e^{\frac{\mu}{ab}}\right\}.
\end{align*}

The stability of the positive coexistence fixed point ($E_{++}$) can be determined using the Jury stability criteria. Since the resulting conditions are lengthy and not explicitly required for our work, we do not state them here. 

\subsection{Existence of Coexistence Fixed Point}\label{ap:cofp2}
The following theorem provides a sufficient condition for the existence of a strictly positive coexistence fixed point.

\begin{theorem}
Model~\eqref{system2noinfect} has a strictly positive coexistence fixed point if
\begin{align*}
\lambda > e^{\frac{\mu}{ab}}.
\end{align*}
\end{theorem}

\begin{proof}
At a fixed point, the equations for $E$ and $L$ satisfy
\begin{align*}
E_{\infty} &= \lambda E_{\infty}e^{-\mu E_{\infty}}e^{-aL_{\infty}},\\
L_{\infty} &= b\lambda E_{\infty}e^{-\mu E_{\infty}}
\left(1-e^{-aL_{\infty}}\right).
\end{align*}
For a strictly positive fixed point, \(E_{\infty}>0\) and \(L_{\infty}>0\). Dividing the first equation by \(E_{\infty}\) and solving for \(E_{\infty}\) gives
\begin{align*}
E_{\infty}
=\frac{\ln(\lambda)-aL_{\infty}}{\mu}.
\end{align*}
Similarly, using the first fixed point equation in the second equation gives
\begin{align*}
L_{\infty}
=bE_{\infty}\left(e^{aL_{\infty}}-1\right),
\end{align*}
and hence
\begin{align*}
E_{\infty}
=\frac{L_{\infty}}{b\left(e^{aL_{\infty}}-1\right)}.
\end{align*}

Therefore, a strictly positive coexistence fixed point corresponds to a positive solution \(L_{\infty}\) of
\begin{align*}
    F(L)
    =\frac{L}{b\left(e^{aL}-1\right)}
    -\frac{\ln(\lambda)-aL}{\mu}
    =0.
\end{align*}

The function $F$ is continuous for $L>0$. As $L\to0^+$,
\begin{align*}
    \lim_{L\to0^+}F(L)
    &=\frac{1}{ab}-\frac{\ln(\lambda)}{\mu}.
\end{align*}
Thus,
\begin{align*}
    \lim_{L\to0^+}F(L)<0
    \quad\Longleftrightarrow\quad
    \lambda>e^{\frac{\mu}{ab}}.
\end{align*}
On the other hand,
\begin{align*}
    \lim_{L\to\infty}F(L)
    =\infty.
\end{align*}

Therefore, if \(\lambda>e^{\frac{\mu}{ab}}\), then \(F(L)\) is negative for sufficiently small \(L>0\) and positive for sufficiently large \(L\). By the intermediate value theorem, there exists some \(L_{\infty}\in(0,\infty)\) such that
\[
    F(L_{\infty})=0.
\]
The corresponding value
\[
    E_{\infty}
    =\frac{L_{\infty}}{b\left(e^{aL_{\infty}}-1\right)}
\]
is also strictly positive. Hence, Model~\eqref{system2noinfect} has a strictly positive coexistence fixed point.
\end{proof}

\subsection{Existence of a Host-Only 2-Cycle}\label{ap:twocycle2}
When $L(t)=0$, the host dynamics reduce to the Ricker map
\begin{align*}
    E(t+1)=\lambda E(t)e^{-\mu E(t)}.
\end{align*}
The following result determines conditions for the existence of a nontrivial host-only 2-cycle.
\begin{theorem}
    Model~\eqref{system2noinfect} has a nontrivial host-only 2-cycle if and only if 
    \begin{align*}
        \lambda>e^2.
    \end{align*}
\end{theorem}
\begin{proof}
    A point belonging to a host-only 2-cycle must be a fixed point of the second iterate of the host-only map. Thus,
    \begin{align*}
        E=\lambda^2Ee^{-\mu E}e^{-\mu \lambda Ee^{-\mu E}}.
    \end{align*}
    For convenience, define $x=\mu E$. Then the two-step fixed-point equation becomes
    \begin{align*}
        x=\lambda^2 xe^{-x}e^{-\lambda x e^{-x}}.
    \end{align*}
    For $x>0$, we may divide by $x$ and take logarithms to obtain
    \begin{align*}
        0=2\ln(\lambda)-x-\lambda xe^{-x}.
    \end{align*}
    Define $g(x)=2\ln(\lambda)-x-\lambda xe^{-x}$. The positive fixed point of the original host-only map is $x_e = \ln(\lambda)$ and it is also a root of $g$ since $g(\ln(\lambda))=0$.

    Thus, a nontrivial positive 2-cycle exists when $g$ has a second positive root. 

    We first consider the case $\lambda>1$ for which
    \begin{align*}
        g(0)=2\ln(\lambda)>0
    \end{align*}
    and 
    \begin{align*}
        \lim_{x \to \infty} g(x) = -\infty.
    \end{align*}

    Furthermore, 
    \begin{align*}
        g'(x) = -1-\lambda e^{-x}+\lambda x e^{-x}.
    \end{align*}
    At the positive fixed point $x_e=\ln(\lambda))$, 
    \begin{align*}
        g'(x_e) &= -1-\lambda e^{-\ln(\lambda)}+\lambda (\ln(\lambda))e^{-\ln(\lambda)}\\
        &= -1-1+\ln(\lambda)\\
        &= \ln(\lambda)-2.
    \end{align*}
    Hence, $g'(x_e)>0$ if and only if $\lambda >e^2$. 

    If $\lambda>e^2$, then $g'(x_e)>0$, so $g$ is negative immediately to the left of $x_e$. Since $g(0)>0$ and $g$ is continuous, there must exist some $x_\infty \in (0,x_e)$ such that $g(x_\infty)=0$. Because $x_\infty \neq x_e$, this corresponds to a nontrivial positive 2-cycle. 

    If instead $1<\lambda \leq e^2$, for $x>1$ we have
    \begin{align*}
        g'(x) = -1-\lambda e^{-x}+\lambda xe^{-x} \leq -1+\lambda e^{-2} \leq 0.
    \end{align*}
    For $0<x\leq 1$, we also have that $g'(x)<0$. Hence, $g$ is strictly decreasing for all $x>0$. Since $g(x_e)=0$, the only positive root of $g$ is $x_e$. Therefore, a nontrivial host-only 2-cycle exists if and only if $\lambda>e^2$.
\end{proof}
\end{appendices}

\bibliography{refs}

\end{document}